\documentclass[aps,prd,twocolumn]{revtex4}
\usepackage[dvips]{epsfig,graphics}
\usepackage{graphicx}

\usepackage{subfigure}
\usepackage{amsmath}

 \newlength{\wth}
 \newcommand{\twographs}[2]{%
 \unitlength=1.1in
 \begin{picture}(5.8,2.6)(-0.3,0)
 \put(0,0){\epsfig{file=#1.eps, width=9cm}}
 \put(2.7,0){\epsfig{file=#2.eps, width=9cm}}
 \put(0.2,2.1){(a)}
 \put(2.7,2.1){(b)}
 \end{picture}
}

\newcommand{\twographschris}[2]{%
 \unitlength=1.1in
 \begin{picture}(6.,2.3)
 \put(0.2,0){\epsfig{file=#1, width=0.7\wth}}
 \put(3.2,0){\epsfig{file=#2, width=0.7\wth}}
 \put(0,2.0){(a)}
 \put(3.0,2.0){(b)}
 \end{picture}
}

\newcommand{\twographst}[2]{%
 \unitlength=1.1in
 \begin{picture}(5.8,2.6)(0.5,0)
 \put(0,0){\epsfig{file=#1, width=\wth}}
 \put(0.9,0.46){\epsfig{file=#12, width=0.678 \wth}}
 \put(2.7,0){\epsfig{file=#2, width=\wth}}
 \put(3.6,0.46){\epsfig{file=#22, width=0.678 \wth}}
 \put(0.5,2.1){(a)}
 \put(3.2,2.1){(b)}
 \end{picture}
}

\newcommand{\fourgraphs}[4]{%
\unitlength=1in
\begin{picture}(8.7,4.7)(-0.4,0.3)
\put(2.7,2.4){\epsfig{file=#2.eps, width=\wth}}
\put(3.69,2.905){\epsfig{file=#22.eps, width=0.678 \wth}}
\put(-0.5,0.1){\epsfig{file=#3.eps, width=\wth}}
\put(0.49,0.605){\epsfig{file=#32.eps, width=0.678 \wth}}
\put(0.1,4.7){(a)}
\put(-0.5,2.4){\epsfig{file=#1.eps, width=\wth}}
\put(0.49,2.905){\epsfig{file=#12.eps, width=0.678 \wth}}
\put(3.3,4.7){(b)}
\put(0.1,2.3){(c)}
\put(3.3,2.3){(d)}
\put(2.7,0.1){\epsfig{file=#4.eps, width=\wth}}
\put(3.69,0.605){\epsfig{file=#42.eps, width=0.678 \wth}}
\end{picture}}

\newcommand{\sixgraphsmod}[6]{%
\unitlength=1in
\begin{picture}(8.7,6.9)(-0.4,0.3)
\put(-0.5,0.){\epsfig{file=#5.eps, width=10cm}}
\put(0.49,0.5){\epsfig{file=#52.eps, width=6.8cm}}
\put(-0.5,4.78){\epsfig{file=#1.eps, width=10cm}}
\put(0.49,5.3){\epsfig{file=#12.eps, width=6.8cm}}
\put(0.1,6.9){(a)}
\put(2.7,4.8){\epsfig{file=#2.eps, width=10cm}}
\put(3.69,5.3){\epsfig{file=#22.eps, width=6.8cm}}
\put(0.1,4.5){(c)}
\put(-0.5,2.3){\epsfig{file=#3.eps, width=10cm}}
\put(0.49,2.8){\epsfig{file=#32.eps, width=6.8cm}}
\put(3.3,4.5){(d)}
\put(2.7,2.3){\epsfig{file=#4.eps, width=10cm}}
\put(3.69,2.8){\epsfig{file=#42.eps, width=6.8cm}}
\put(0.1,2.2){(e)}
\put(3.3,2.2){(f)}
\put(2.7,0.){\epsfig{file=#6.eps, width=10cm}}
\put(3.69,0.5){\epsfig{file=#62.eps, width=6.8cm}}
\put(3.3,6.9){(b)}
\end{picture}}

\begin{document}

\title{Multi-Dimensional mSUGRA Likelihood Maps\footnote{Preprint number: DAMTP-2005-64}}

\author{B.C. Allanach}
\affiliation{DAMTP, CMS, Wilberforce Road,
Cambridge, CB3 0WA, United Kingdom}

\author{C.G. Lester}
\affiliation{Cavendish Laboratory. Madingley Road.
Cambridge CB3 0HE, United Kingdom}

\begin{abstract}
  We calculate the likelihood map in the full 7 dimensional parameter
  space  of the minimal supersymmetric standard model (MSSM) assuming universal
  boundary conditions on the supersymmetry breaking terms. Simultaneous
  variations of $m_0$, $A_0$, $M_{1/2}$, $\tan \beta$, $m_t$, $m_b$ and
  $\alpha_s(M_Z)$ are applied using a Markov chain Monte Carlo algorithm.
  We use measurements of $b \rightarrow s \gamma$, $(g-2)_{\mu}$ and
  $\Omega_{DM} h^2$ in order to constrain the model. We present likelihood
  distributions for some of the sparticle 
  masses, for the branching ratio of
  $B_s^0 \rightarrow \mu^+ \mu^-$ and for $m_{\tilde \tau}-m_{\chi_1^0}$. 
  An upper limit of 2$\times 10^{-8}$ on this branching ratio might be achieved
  at the Tevatron, and would rule out 29$\%$ of the currently allowed
  likelihood. If one allows for non thermal-neutralino components of dark
  matter, this fraction becomes 35$\%$. 
  The mass ordering allows the important cascade decay
  ${\tilde q}_L \rightarrow {\chi_2^0} \rightarrow {\tilde l}_R
  \rightarrow {\chi_1^0}$ with a likelihood of 24$\pm 4\%$. The stop
  coannihilation region is highly disfavoured, whereas the light Higgs region 
  is marginally disfavoured.
\end{abstract}

\maketitle

\section{Introduction}
Weak-scale supersymmetry provides a well-documented solution to the technical
hierarchy problem~\cite{Martin:1997ns}, which is particularly difficult to
solve in a perturbatively calculable model. 
Specialising to a minimal extension of the Standard Model, the MSSM, one can
provide a weakly interacting massive particle dark matter candidate,
provided $R-$parity is respected by the model. 
Examples of dark matter candidates are the
gravitino~\cite{Ellis:2003dn,Roszkowski:2004jd}, the axino~\cite{Covi:2004rb}
and the lightest neutralino~\cite{Ellis:1983ew}, the subject of much
recent investigation~\cite{Baer:2002gm,Ellis:2003cw,Battaglia:2003ab,Ellis:2003si,Gomez:2004ek,Ellis:2004tc}. 
The general MSSM is rather complicated due to the large number of free
parameters in the supersymmetric (SUSY) breaking sector. 
However, the observed rareness of flavour changing neutral currents (FCNCs)
suggests  
that the vast majority of parameter space for general SUSY breaking terms is
ruled out. 
Particular patterns of SUSY breaking parameters can postdict small enough
FCNCs: for instance flavour universality. 
One highly studied subset of such terms is that of mSUGRA, often called
the Constrained Minimal Supersymmetric Standard Model (CMSSM).
In mSUGRA, at some high energy scale (typically taken to be the scale of
unification of electroweak gauge couplings), all of the SUSY breaking scalar
mass terms are assumed to be equal to $m_0$, the scalar trilinear terms 
are set to $A_0$ and the gaugino masses are set equal ($M_{1/2}$). 
These are indeed strong assumptions, but they have several advantages for
phenomenological analysis as the number of independent SUSY breaking
parameters is much 
reduced. Indeed, assuming that the MSSM is the correct model, the initial
data from the Tevatron or Large Hadron Collider are likely to contain only a
few relevant 
observables and so one may be able to fit against a simple SUSY breaking model 
as an example~\cite{Armstrong:1994it}. As the data become more accurate and
additional 
relevant observables are measured, the lack of a good fit would propel 
extensions 
of the simple model. One may then start to consider
patterns of non-universality, for instance. 
For the rest of this paper though, given the lack of data to the contrary,
we will assume mSUGRA\@.
Aside from the universal soft terms $m_0, A_0, M_{1/2}$,
other non-standard model mSUGRA input parameters are taken to be $\tan
\beta$, the ratio of the two Higgs vacuum expectation values, and
the sign of $\mu$ (a parameter that appears in the Higgs potential of the
MSSM).  

When combined with large scale structure data,
the Wilkinson microwave ani\-sot\-ropy probe (WMAP)~\cite{Spergel:2003cb,Bennett:2003bz} has placed 
stringent constraints upon the dark matter relic density $\Omega_{DM} h^2$.
A common assumption, which we will adhere to here, is
that the neutralino makes up the entire cold dark matter relic density.
The prediction of the relic density of dark matter in the MSSM depends
crucially upon annihilation cross-sections, since in the early universe 
SUSY particles will annihilate in the thermal bath.\ 
Regions of 
mSUGRA that are compatible with the WMAP constraint often predict some of the
following annihilation channels ~\cite{Allanach:2004xn}:
\begin{itemize}
\item
Stau (${\tilde \tau}$) co-annihilation~\cite{Griest:1990kh} at small $m_0$
where the lightest stau is quasi-degener\-ate with the lightest neutralino
($\chi_1^0$).  
\item
Pseudoscalar Higgs ($A^0$) funnel region at large $\tan \beta > 45$ where
two neutralinos annihilate through an $s$-channel $A^0$
resonance~\cite{Drees:1992am,Arnowitt:1993mg}. 
\item
Light CP-even Higgs ($h^0$) region at low $M_{1/2}$ where
two neutralinos annihilate through an $s$-channel $h^0$
resonance~\cite{Drees:1992am,Djouadi:2005dz}. 
\item
Focus point~\cite{Feng:1999mn,Feng:1999zg,Feng:2000gh} at large $m_0$ where a
significant 
Higgsino component leads to efficient neutralino annihilation into gauge boson
pairs.
\item
Stop co-annihilation~\cite{Boehm:9911,arnie,Ellis:2001nx} at large $(-A_0)$,
where the lightest 
stop is close in mass to the lightest neutralino.
\end{itemize}
Many pre-WMAP analyses focused on the so-called bulk
region~\cite{Drees:1992am}. The 
bulk region is continuously connected to the
stau co-annihilation region at low $m_0$ and $M_{1/2}$.
There are two reasons why the bulk region has shrunk in size when
one takes the current constraints into account: 
the WMAP constraint upon the relic density has ruled much of the region out,
and the new low value of the top mass mean that the MSSM Higgs mass
predictions are sometimes too low for low $m_0$ and $M_{1/2}$
and are ruled out by LEP2 constraints. 
The (now reduced) bulk region will make an implicit appearance in our
results, and we will comment upon this fact later.

Several
authors~\cite{Brhlik:2000dm,Drees:2000he,Polesello:2004qy,Bambade:2004tq,Battaglia:2003ab,Allanach:2004jh,Moroi:2005nc}
have asked how the annihilation cross-section can be 
constrained 
by collider measurements in order to provide a more solid prediction of the
relic density. This would then be fed into a cosmological model in
order to predict  $\Omega_{DM} h^2$ for comparison
with the value derived from cosmological observation, allowing a test of
cosmological assumptions (and the assumption that there is only one component
of cold dark matter). Of course, colliders could not unambiguously identify
the lightest 
observed SUSY particle as the dark matter since it could always decay
unobserved outside the detector. 
It would therefore be interesting to combine collider
information with  that derived from a possible future direct
detection~\cite{Bourjaily:2005ax} of dark matter, providing corroboration and additional
empirical information. 
Before such observations are made, however, we may  ask how well current data
constrain models of new physics. 

This question has been addressed many times for mSUGRA by using the dark
matter constraint. Most of the analyses (see, for
example~\cite{Roszkowski:2001sb,Baer:2002gm,Baer:2002fv,Baer:2003yh,Ellis:2003cw,Chattopadhyay:2003xi,Lahanas:2003yz,Belanger:2004hk})   
fix all but two parameters and examine constraints upon the remaining 2
dimensional (2d) 
slice of parameter space. The dark matter relic density constraint is the most
limiting, but the branching ratio of the decay $b \rightarrow s \gamma$ 
and the anomalous magnetic moment of the muon $(g-2)_\mu$ also rule part of
the parameter space out. Recent upper bounds from the Tevatron
  experiments on the branching ratio $B_s 
\rightarrow \mu^+ \mu^-$~\cite{Herndon:2004tk} have the potential to 
restrict mSUGRA in the future, but the analysis of ref.~\cite{Ellis:2005sc}
shows that the resulting constraints currently subsumed within other
constraints. In the above analyses,
limits are typically imposed separately, each to some 
prescribed confidence level. Such analyses have the advantage of being
quite transparent: it is fairly easy to see which constraint rules out which
part of parameter space. However, they have the disadvantages of not properly
describing the combination of likelihoods coming from different experimental
constraints and of having to assume {\em ad hoc}\/ values for several input
parameters. In particular, as well as the soft SUSY breaking input parameters,
the bottom mass $m_b$, the strong structure
constant $\alpha_S(M_Z)$ and the top mass $m_t$ can all have a strong effect on 
mSUGRA predictions.
A large random scan of flavour diagonal MSSM space involving $10^5$ points
that pass various prescribed constraints was presented in
ref.\cite{Profumo:2004at}, 
however the sampling of the 20d parameter space was necessarily
sparse. The analysis is also subject to the limitation that likelihoods have
not been combined; instead the measurements have been used as cuts to discard
points. 
In ref.~\cite{Ellis:2003si}, the likelihood from the observables is
calculated, properly combining different constraints, but again 1d and 2d
slices through parameter space were taken.
Of course the time taken to efficiently sample from a likelihood distribution
using the naive method (a scan) scales like a power law with respect to
the number of parameters, meaning that in practice even a high
resolution 3d scan is difficult. By parameterising lines in 2d that are
consistent with 
the WMAP dark matter constraint and scanning in two other parameters, the
analysis of 
ref.~\cite{Ellis:2004tc} 
calculates the $\chi^2$ statistic
for the 2d part of a 3d parameter space which is
consistent with the WMAP constraint on the dark matter relic density.
The predicted value of $\Omega_{DM} h^2$ is not combined in the
  $\chi^2$ with the other observables
for this analysis, and the parameter $\tan \beta$ must be fixed. 
As the authors note~\cite{Ellis:2004tc}, parts of the scan were
sparse. In Ref.~\cite{Stark:2005mp}, a scan was performed which included
variations of $A_0$ and $\tan \beta$ as well as other mSUGRA parameters. It is
clear from this paper that the WMAP allowed region 
(expressed in the $M_{1/2}$-$m_0$ plane) becomes much larger from the $A_0$
variations. No likelihood distribution was given. 

Baltz and Gondolo~\cite{Baltz:2004aw} demonstrated that a Markov chain Monte
Carlo (MCMC) algorithm efficiently samples from the mSUGRA parameter space,
rendering 4d scans in $m_0, A_0, M_{1/2}, \tan \beta$
feasible. However, they were interested in  
which parts of parameter space are compatible with the WMAP measurement of
$\Omega_{DM} h^2$ and what the prospects are for direct detection there, not
in the likelihood distribution. In order
to increase the efficiency of their parameter sampling, they changed
the simple ``Metropolis-Hastings'' MCMC algorithm in order to achieve a better
efficiency. As the authors state in their
conclusions, this has the consequence that
caution must be exercised when trying to interpret their
results as a likelihood distribution.
Indeed, we will show in a toy model that changes to the MCMC algorithm like
the ones that Baltz and Gondolo made can alter the sampling
from a distribution. 

It is our purpose here to utilise the MCMC algorithm in such a way as to
reliably calculate the combined likelihood of mSUGRA in the full
dimensionality of its parameter space, thereby extending the previous
studies. We will then be 
able to infer what is known about the multi-dimensional
parameter space, including important variations of the SM quantities.
These results will have implications for collider searches and  rare decays.

In section~\ref{sec:mcmc}, we briefly review the MCMC algorithm. 
We present the implementation used in the present paper to 
calculate the likelihood maps of mSUGRA parameter space and then demonstrate
that the results are convergent using a particular statistical test.
In section~\ref{sec:Lmaps}, we present the likelihood distributions and
derived quantities of the 7d
mSUGRA parameter space. In section~\ref{sec:uncertainties}, we illustrate the
effects of theoretical uncertainties in the sparticle spectrum calculation and
in section~\ref{sec:other}, possible effects from allowing an additional non
thermal-neutralino component to the relic density are explored.
A summary and conclusions are presented in  section~\ref{sec:conc}.
In appendix~\ref{sec:badsampling}, we 
demonstrate with two different toy models that the algorithm
used by Baltz and Gondolo may not provide a sampling proportional to the
likelihood of the parameter space.

\section{Implementation of the MCMC Algorithm\label{sec:mcmc}}

\subsection{Likelihoods}
Some readers might be unfamiliar with the use of statistics in this paper, and
so we include some comments on how to interpret them. The 
likelihood is {\em not}
dependent upon any priors. 
The likelihood ${\mathcal L} \equiv p(d|m)$ is the probability density
function (pdf) of reproducing
data $d$ 
assuming some mSUGRA model $m$. In $p(d|m)$, the model $m$ is specified by the
mSUGRA input parameters and so $p(d|m)$ has a dependence upon them.
$p(d|m)$ is related to the pdf of the model being the one chosen by nature,
given the data, by an application of Bayes' theorem:
\begin{equation}
p(m|d) = p(d|m) \frac{ p(m) }{ p(d) }, \label{bayes}
\end{equation}
where $p(m)$ is the probability of the model being correct (the {\em prior})
and $p(d)$ is the total probability of the data being reproduced, integrating
over all possible models. $p(d)$ is
practically impossible to estimate, so we cannot get the quantity that one
really wishes to estimate, $p(m|d)$. 
However, we may compare the relative probabilities of two different models
$m_1$ and $m_2$ (corresponding here to different points of mSUGRA space) by
applying 
Eq.~\ref{bayes} for each model, implying that
\begin{equation}
\frac{p(m_1|d)}{p(m_2|d)} = \frac{ p(d|m_1) p(m_1) }{p(d|m_2) p(m_2)}.
\end{equation}
We note here the appearance of the infamous prior distributions $p(m_1),
p(m_2)$. If one assumes that the ratio of these two priors is one (that no
region of parameter space is more likely than any other), one may interpret
the likelihood ratio of two different points in mSUGRA space as the ratio of
probabilities of the models, given the data. 
In this paper however, we provide likelihood distributions. If the reader
prefers a different ratio of priors to one, they must convolute the likelihood
density we give with their preferred ratio of pdfs. 

\subsection{The MCMC Algorithm}

We now briefly review the Metropolis MCMC algorithm, but for a more
thorough explanation, see refs.~\cite{Baltz:2004aw,MacKay}.
Other adaptive scanning algorithms have recently been
suggested in the context of high energy
physics~\cite{Allanach:2004my,Brein:2004kh} but (although they can be very
useful for 
other purposes) they do not yield a likelihood distribution. 
A Markov chain consists of a list of parameter points (${\mathbf x}^{(t)}$)
and associated 
likelihoods (${\mathcal L}^{(t)} \equiv {\mathcal L}({\mathbf x}^{(t)})$). Here
$t$ labels 
the link number in the chain.
Given some point at the end of the Markov chain (${\mathbf x}^{(t)}$),
the Metropolis-Hastings algorithm involves randomly picking another potential
point (${\mathbf x}^{(t+1)}$) (typically in the vicinity of ${\mathbf
  x}^{(t)}$)
 using
some proposal pdf $Q({\mathbf x};{\mathbf x}^{(t)})$. In this paper we will
specialise to the case of symmetric proposal functions, i.e.\ $Q({\mathbf
  x_b}; {\mathbf x_a})=Q({\mathbf x_a}; {\mathbf x_b})$. 
If ${\mathcal L}^{(t+1)}>{\mathcal L}^{(t)}$, the new point is appended onto
the 
chain. Otherwise, the proposed point is accepted with probability
${\mathcal L}^{(t+1)} / {\mathcal L}^{(t)}$ and, if accepted, added to the end
of the   chain. If 
the point ${\mathbf x}^{(t+1)}$ is not accepted, the point ${\mathbf
  x}^{(t)}$ is copied on to the end of the chain instead.

Providing ``detailed balance'' is satisfied, it can be shown~\cite{MacKay}
that the sampling density of points in the chain is proportional to 
the target distribution (in this case, the likelihood) 
as the number of links goes to infinity. 
In the context of this analysis, detailed balance states that for any two
points  ${\mathbf x_a}, {\mathbf x_b}$
\begin{equation}
T({\mathbf x_a} ; {\mathbf x_b}) {\mathcal L}({\mathbf x_b}) = T({\mathbf x_b}
; {\mathbf x_a}) {\mathcal L}({\mathbf x_a}),
\end{equation}
where $T({\mathbf x_b}; {\mathbf x_a}) \equiv Q({\mathbf x_b}; {\mathbf x_a})
\times \mbox{min}(1, {\mathcal L}({\mathbf x_b})/{\mathcal L}({\mathbf x_a}))$ is the probability of
 making a transition from ${\mathbf x_a}$ to ${\mathbf x_b}$ in the case where
 the proposal function 
 is symmetric. 
Thus, the probability of sampling a point ${\mathbf x_a}$ from the likelihood
distribution and then making a 
transition to ${\mathbf x_b}$ 
be equal to the probability of sampling ${\mathbf x_b}$ and making a
transition to ${\mathbf x_a}$.

The Metropolis-Hastings MCMC algorithm is typically much more efficient than a 
straightforward scan for $D>3$; the number of required steps scales roughly
linearly 
with $D$ rather than as a power law. The sampling is in principle independent
of the form of $Q$ as $t \rightarrow \infty$ as long as it is bigger than zero
everywhere. However,
$Q$ must be chosen with some care: since in practice we can only sample a
finite number of points, the choice of the form of $Q$ can determine whether 
the entire parameter space is sampled and how quickly convergence is reached. 

Baltz and Gondolo used a geometrical model for $Q$: choosing a random distance
from the point ${\mathbf x^{(t)}}$ and using a direction that was calculated
  from the positions of previous 
points in the chain. The width of the random radius pdf was calculated
depending upon previous points in the chain in order to increase the
efficiency of the 
calculation, aiming to accept roughly 25$\%$ of potential points. Either of
these changes upset detailed balance and may spoil the sampling. We
demonstrate in Appendix~\ref{sec:badsampling} with toy models
that the width changing modification gives a sampling that is not proportional
to the target density. 

\subsection{Parameter Ranges \label{sec:imp}}

\begin{table}
\caption{Parameter ranges considered. \label{tab:prior}}
\begin{tabular}{|c|c|}
\hline
parameter & range \\ \hline
{sign}$(\mu)$ & +1\ \\
$A_0$ & -2 TeV$-$2 TeV \\
$m_0$ & 60 GeV$-$2 TeV \\
$M_{1/2}$ & 60 GeV$-$2 TeV \\
$\tan \beta$ & 2$-$60 \\ \hline
\end{tabular}
\end{table}
On general naturalness grounds\footnote{That
  is, to avoid a large cancellation between weak scale SUSY breaking terms in
  order to get a small value of $\mu$.}, we expect $m_0$ and $M_{1/2}$ to not
be too large: less than say, 2 TeV. 
mSUGRA is ruled out by negative results in sparticle searches for
$m_0<60$ GeV or $M_{1/2}<60$ GeV. $\mu>0$ is favoured by the measurement of
the anomalous magnetic moment of the muon.
$\tan \beta$ is bounded from below by
negative searches at LEP2 for $h^0$ (and perturbativity of the top Yukawa
coupling) and from above by perturbativity of the Yukawa couplings up to the
unification scale. 

Upper bounds upon $m_0$ can 
exclude the focus point region, which, in our calculation,
is at much higher values of $m_0\sim O(8)$ TeV.
It has been
argued that a quantitative measure of fine-tuning in the focus point region is
not too large~\cite{Feng:1999mn,Feng:1999zg}, however the
fine-tuning of the top quark Yukawa coupling is
enormous~\cite{Allanach:2000ii}. 
This makes the focus-point regime practically impossible to reliably
calculate starting from mSUGRA inputs. Tiny higher order effects in
the top Yukawa coupling strongly
change the position of the focus point regime in mSUGRA parameter space.
In ref.~\cite{Allanach:2000ii}, it is demonstrated that the focus point
regime 
moves in the $m_0$ direction by several TeV depending on how exactly the
highest order top-quark Yukawa radiative corrections are calculated.
Because the calculation cannot be controlled with the current state-of-the-art
calculations of the top Yukawa coupling, we will
exclude the focus point regime from this analysis by placing an appropriate
upper bound upon $m_0$.
Here, we restrict the parameter space to that shown in Table~\ref{tab:prior}.

\subsection{Observables and constraints}

\begin{table}
\caption{Lower bounds applied to sparticle mass predictions in GeV.\label{tab:constraints}}
\begin{tabular}{|cc|cc|cc|cc|}
\hline
$m_{\chi_1^0}$ & 37  & $m_{\chi^\pm_1}$ & 67.7 & 
$m_{\tilde g}$ & 195 &
$m_{{\tilde \tau}_1}$ & 76 \\ $m_{{\tilde l}_R}$ & 88 & $m_{{\tilde t}_1}$ &
  86.4 &
$m_{{\tilde b}_1}$ & 91 & $m_{{\tilde q}_R}$ & 250 \\
$m_{{\tilde \nu}_{e,\mu}}$ & 43.1 & & & & & & \\
\hline
\end{tabular}
\end{table}
We calculate the MSSM spectrum from mSUGRA
parameters, by using the program {\tt SOFTSUSY1.9.2}~\cite{Allanach:2001kg}.
Ideally we would like to include data from negative search results from
collider data within a combined likelihood. 
Unfortunately it is difficult to obtain the data in such a form and so
instead, we assign a 
zero likelihood to any point for which at least one of the
constraints~\cite{PDBook} in Table~\ref{tab:constraints} is not satisfied.
We also implement a parameterisation\footnote{Developed by P Slavich for
  Ref.~\protect\cite{Allanach:2004rh}.} of the 95$\%$
confidence level 
limits~\cite{Barate:2003sz} on $m_h(g_{hZZ}/g_{hZZ}^{SM})$, where
$g_{hZZ}/g_{hZZ}^{SM}$ is the ratio of the MSSM higgs coupling to two $Z^0$
bosons to the equivalent Standard Model coupling. In order to take a 3 GeV
uncertainty on  
the mSUGRA prediction of $m_h$ into account, we
add 3 GeV~\cite{Degrassi:2002fi,Allanach:2004rh}
to the $m_{h^0}$ value that is used in the parameterisation. In the MSSM,
$g_{hZZ}/g_{hZZ}^{SM}=\sin(\beta - \alpha)$ and
in practice, it
is easier to apply limits in terms of the {\em inverse} \/parameterisation
$\sin^2 (\beta - \alpha)(m_{h^0})$ as shown in Table~\ref{tab:par}.

\begin{table*}
\caption{Parameterisation of 95$\%$ condidence level LEP2 Higgs limits on the
  $m_{h^0}$-$\sin^2(\beta-\alpha)$ plane. All points with $m_{h^0}<90$ GeV are
  ruled out. \label{tab:par}}
\begin{tabular}{|c|c|}
\hline $m_{h^0}/$GeV range & upper bound on $\sin^2 (\beta -
  \alpha)$ \\ \hline
90-99 & -6.1979 + 0.12313 $m_{h^0}/$GeV - 0.00058411 $(m_{h^0}/$GeV$)^2$\\
99-104 & 35.73 - 0.69747 $m_{h^0}/$GeV + 0.0034266 $(m_{h^0}/$GeV$)^2$\\
104-109.5 & 21.379 - 0.403 $m_{h^0}/$GeV + 0.0019211 $(m_{h^0}/GeV)^2$ \\
109.5-114.4 & 1/(60.081 - 0.51624 $m_{h^0}/$GeV) \\ \hline
\end{tabular}
\end{table*} 
The spectrum is transferred via the SUSY Les 
Houches Accord~\cite{Skands:2003cj} to the computer program {\tt
  micrOMEGAs1.3.5}~\cite{Belanger:2001fz,Belanger:2004yn} in order to
calculate several quantities used to calculate the likelihood of a
parameter point. 
We will use six measurements in order to construct the final likelihood of any
given point of parameter space. As mentioned in the introduction, we make the 
assumption that the neutralino makes up the entire cold dark matter relic
density as constrained by WMAP:
\begin{equation}
\Omega_{DM} h^2 = 0.1126^{+0.0081}_{-0.0091}. \label{omConst}
\end{equation}
The anomalous magnetic moment of the muon has been
 measured~\cite{Bennett:2004pv} to be higher than the Standard Model
 prediction~\cite{Passera:2004bj,deTroconiz:2004tr}. The experimental
 measurement is so precise that the comparison is limited by theoretical
 uncertainties in the Standard Model prediction. Following
 Ref.~\cite{Allanach:2005yq}, we constrain any new physics contribution to 
 be
\begin{equation}
\delta \frac{(g-2)_\mu}{2} = 19.0 \pm 8.4 \times 10^{-10}. 
\end{equation}
Adding theoretical errors~\cite{Gambino:2004mv} to measurement
errors~\cite{hfag} in quadrature 
for the branching ratio for the decay $b \rightarrow s \gamma$,
one obtains the empirically derived constraint
\begin{equation}
BR(b \rightarrow s \gamma) = 3.52 \pm 0.42 \times 10^{-4}.
\end{equation}

The Standard Model inputs' measurements also contribute to the
likelihood. We take these to be~\cite{PDBook}, for the running bottom quark
mass in the modified minimal subtraction scheme,
\begin{equation}
m_b (m_b)^{\overline{MS}} = 4.2 \pm 0.2 \mbox{~GeV},
\end{equation}
for the pole mass of the top quark\footnote{For an analysis with 
$m_t=174.3 \pm 3.4$ GeV, see the original version of this paper on the
  hep-ph/ electronic archive.}~\cite{Group:2005cc},  
\begin{equation}
m_t = 172.7 \pm 2.9 \mbox{~GeV},
\end{equation}
and for the strong coupling constant in the modified minimal subtraction
scheme at $M_Z$
\begin{equation}
\alpha_s (M_Z)^{\overline{MS}} = 0.1187 \pm 0.002.
\end{equation}
A prediction $p_i$ of one of these quantities, where
\begin{eqnarray}
i&\equiv& \{
\alpha_s(M_Z)^{\overline{MS}}, m_t, m_b(m_b)^{\overline{MS}},
(g-2)_\mu/2, \nonumber \\
&& \mbox{BR}(b \rightarrow s \gamma), \Omega_{DM} h^2\}
\end{eqnarray}
 with
measurement $m_i\pm s_i$  yields a log likelihood 
\begin{equation}
\ln {\mathcal L}_{i} = - \frac{(m_i-p_i)^2}{2s^2_i} - \frac{1}{2} \ln (2 \pi) -
\ln s_i, 
\label{lnL}
\end{equation}
assuming the usual Gaussian errors. 
Note that since Eq.~\ref{omConst}
has asymmetric errors, $s_{\Omega_{DM} h^2}=0.0081(0.0091)$ if our prediction
is higher (lower) than the observed central value. 
To form the combined likelihood, one 
takes $\ln {\mathcal L}^{tot} = \sum_{i=1} \ln {\mathcal L}_i$,
corresponding to the combination of independent Gaussian likelihoods.
In practice, we will ignore the normalisation
constants $-\frac{1}{2}\ln(2 \pi) - \ln s_i$, since the likelihood
distribution has an arbitrary normalisation anyway. 

We take 
the proposal function to be the product of Gaussian distributions along
each dimension $k=1,2,\ldots,D$ centred on the location of the current point 
along that dimension, i.e.\ $x^{(t)}_k$: 
\begin{equation}
Q({\mathbf x}^{(t+1)}; {\mathbf x}^{(t)}) = 
  \prod_{k=1}^D \frac{1}{\sqrt{2 \pi} l_k} e^{-(x_k^{(t+1)} -x_k^{(t)} )^2
  / 2 l_k^2},
\end{equation}
where $l_k$ denotes the width of the distribution along direction $k$. 
By trial and error we find that using values of $l_k$ that are equal to the
parameter range of dimension $k$ given in Table~\ref{tab:prior}
divided by 25 works well. 
For the Standard Model inputs, we choose $l_k = 8 \sigma_k / 25$. 

In order to start the chain we follow the following procedure, which finds a
point at random in parameter space that is not a terrible fit to the data.
We pick some ${\mathbf y}^{(0)}$ at random in the mSUGRA parameter
space using a flat distribution for its pdf. 
The Markov chain for ${\mathbf y}$ is evolved a sufficient number of steps
($t$) such that $\ln{\mathcal L }({\mathbf y}^{(t)}) > -5$, i.e.\ the initial
chain 
has found a reasonable fit. We then set ${\mathbf x}^{(0)}={\mathbf y}^{(t)}$,
continuing the Markov chain in ${\mathbf x}$ and discarding the ``burn-in''
chain ${\mathbf y}$. The reasonable-fit point is typically found long before
2000 iterations of the Markov chain.

\subsection{Convergence}

In order that likelihood distributions calculated in this paper be considered
reliable, it is important to check convergence of the MCMC\@. This is done by
running 9 independent Markov chains, each with random starting positions as
described above.  
The starting positions are chosen in the ranges presented in
Table~\ref{tab:prior} 
with a flat pdf, since it is important for the convergence measure that the
initial values be over distributed compared to the likelihood function one
samples from. By examining the variance and means of input parameters within the chains
and between the 9 different chains, we will construct a
quantity~\cite{stat}
$\hat R$. $\hat R$ will provide an upper bound on the factor of expected
decrease of 
variance of 1d likelihood distributions
if  the chain were iterated to an infinite number of steps. 
An ${\hat R}$ value can be constructed for scalar quantities that are
associated with a point ${\mathbf x}^{(t)}$. 

The analysis of the $\hat R$ convergence statistic follows Ref.~\cite{stat}
closely. 
We consider $c=1,\ldots,M$ chains ($M=9$ here), each with $N=10^6$ steps. Then
we may define the  average input parameter along direction $k$ for the chain
$c$ and the average amongst the ensemble of chains 
\begin{equation}
[\bar{x}_k]_c = \frac{1}{N} \sum_{t=1}^{N} [{x_k^{(t)}}]_c, \qquad
\bar{x}_k = \frac{1}{M} \sum_{c=1}^{M} [\bar{x}_k]_c,
\end{equation}
respectively.
The variance of chain $c$ along direction $k$ is
\begin{equation}
[V_k]_c = \frac{1}{N-1} \sum_{t=1}^N ([x^{(t)}_k]_c - [\bar{x}_k]_c)^2,
\end{equation}
so that we have the average of the variances within a chain
\begin{equation}
w_k = \frac{1}{M} \sum_{c=1}^M [V_k]_c
\end{equation}
and the variance between chains' averages
\begin{equation}
B_k/N = \frac{1}{M-1} \sum_{c=1}^M ([\bar{x}_k]_c - \bar{x}_k)^2.
\end{equation}
The basic ratio constructed corresponds to
\begin{equation}
R_k = \frac{\frac{N-1}{N} w_k + B_k/N(1 + \frac{1}{M})}{w_k}.
\end{equation}
As long as the initial seed parameters of the Markov chain are
over-distributed, i.e.\ they have larger variance than the likelihood, 
this ratio will be larger than one~\cite{stat} if the chains have not
converged or if they have not had time to explore the entirety of the
parameter space. It tends to one only if both of these conditions are met.
In order to construct $\hat R$, we must take into account the sampling
variability of $[\bar{x}_k]_c$ and $[V_k]_c$.
The variance of chain variances along direction $k$ is estimated to be
\begin{equation}
v_k = \frac{1}{M-1} \sum_{c=1}^M ([V_k]_c - w_k)^2,
\end{equation}
and we must take into account the following estimates of
co-variances between the values of
$[\bar{x}_k]_c$ and $[V_k]_c$:
\begin{eqnarray}
(\sigma_k)_1 &=& - w_k \bar{x}_k^2 + \frac{1}{M}\sum_{c=1}^M [V_k]_c
  [\bar{x}_k]_c^2 , \nonumber \\
(\sigma_k)_2 &=& - w_k
  \bar{x}_k + \frac{1}{M} \sum_{c=1}^M [V_k]_c [\bar{x}_k]_c.
\end{eqnarray}
Defining the total estimated variance of the target distribution along
direction $k$ 
\begin{eqnarray}
{\mathcal V_k} &=& \frac{1}{M}(1 - \frac{1}{N})^2 v_k +
\frac{2(M+1)^2}{M^2(M-1)} 
(B_k/N)^2 +\nonumber \\
&& 
2 \frac{(N-1)(M+1)}{M^2 N} \left( (\sigma_k)_1 - 2 \bar{x}_k 
(\sigma_k)_2
\right),
\end{eqnarray}
where we have degrees of freedom
\begin{equation}
df_k = 2 \frac{\left(R_k w_k + B_k / (NM)\right)^2}{{\mathcal V}_k},
\end{equation}
leading us to the final equation for the estimated reduction in the sampled
variance as $t \rightarrow \infty$:
\begin{equation}
\hat{R}_k = R_k \frac{df_k}{df_k - 2}.
\end{equation}
\begin{figure}
  \epsfig{figure=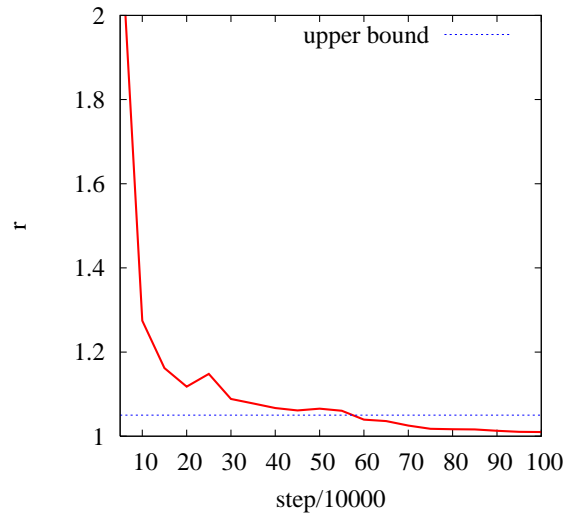,width=10cm}
  \caption{Estimate of potential scale reduction shown as a function of the
    number of Markov
    chain Monte Carlo steps. The upper bound we require for convergence is shown
    as a horizontal line. \label{fig:conv}}
\end{figure}
 

Here,
we define $r \equiv$max$_k \{ \sqrt{\hat R_k}\}$.
Values of
$r<1.05$ are considered to signify convergence and compatibility of the
chains, since we could only hope to decrease the scale of any of the input
parameter distributions by at most $5\%$ by performing further Markov Chain
steps.  
In Fig.~\ref{fig:conv}, we show the quantity $r$ as a function of step
number. $r<1.05$ is met already for 600 000 steps indicating adequate
  convergence, although for the 
  results we present below we we always use the full 9$\times$1 000 000
  sample.

\section{Likelihood Maps\label{sec:Lmaps}}

The number of input parameters exceeds the number of data and
the likelihood shows a rough degeneracy along
directions 
which give iso-lines of $\Omega_{DM} h^2$. The parameters of the best-fit
point of the MCMC do not 
therefore supply us with much information, but the value of the likelihood at
that point is interesting: a very small value would indicate a high $\chi^2$
and therefore a bad fit. 
The best-fit point sampled by the MCMC with 7d input parameter space was 
\begin{eqnarray}
&&m_0=964 \mbox{~GeV},\ M_{1/2}=341 \mbox{~GeV},\ A_0=1394 \mbox{~GeV},
\nonumber \\
&&
m_b(m_b)=4.18 \mbox{~GeV},\ m_t=173.0 \mbox{~GeV},\ \nonumber \\
&& \alpha_s(M_Z)=0.1185,\
\tan \beta = 57.9, 
\end{eqnarray}
leading to predictions of $\delta (g-2)_\mu/2=1.8 \times 10^{-9}$, $BR(b
\rightarrow s 
\gamma)=3.63 \times 10^{-4}$ and $\Omega_{DM} h^2=0.1124$ and corresponding to
a 
combined likelihood (ignoring normalisation constants, as stated above) 
of ${\mathcal L}=0.95$. 
The point is within the $A^0-$pole region, and the centrality of predicted
observables 
gives us confidence that mSUGRA can fit well to current data.
The efficiency of the MCMC algorithm is quite small: only 4.1$\%$. This
reflects the fact that the thickness of WMAP-allowed volume is small, making
it difficult to sample efficiently. 

\begin{figure*}
\begin{center}
\sixgraphsmod{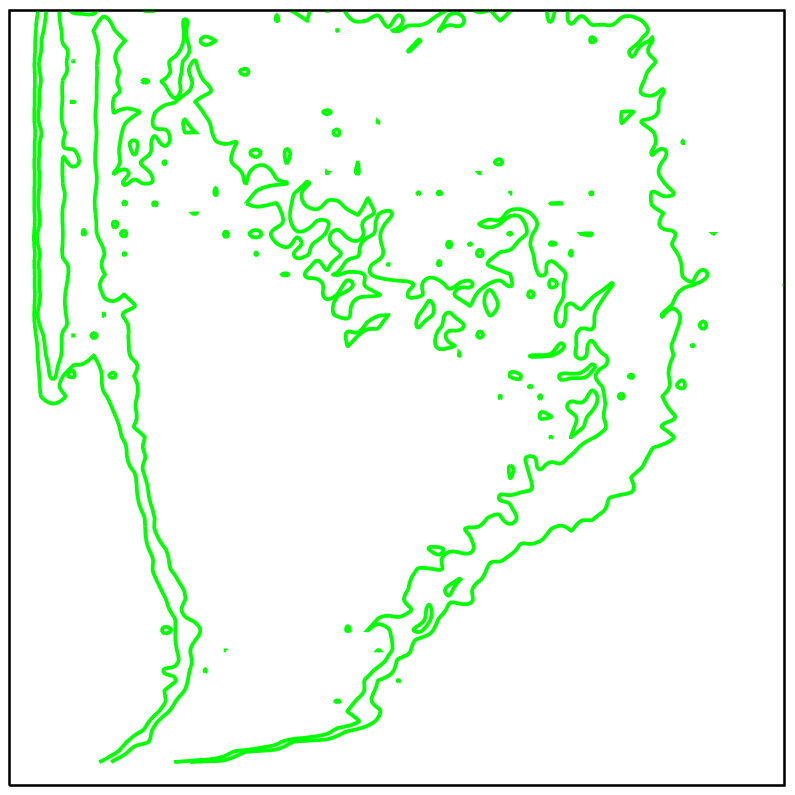}{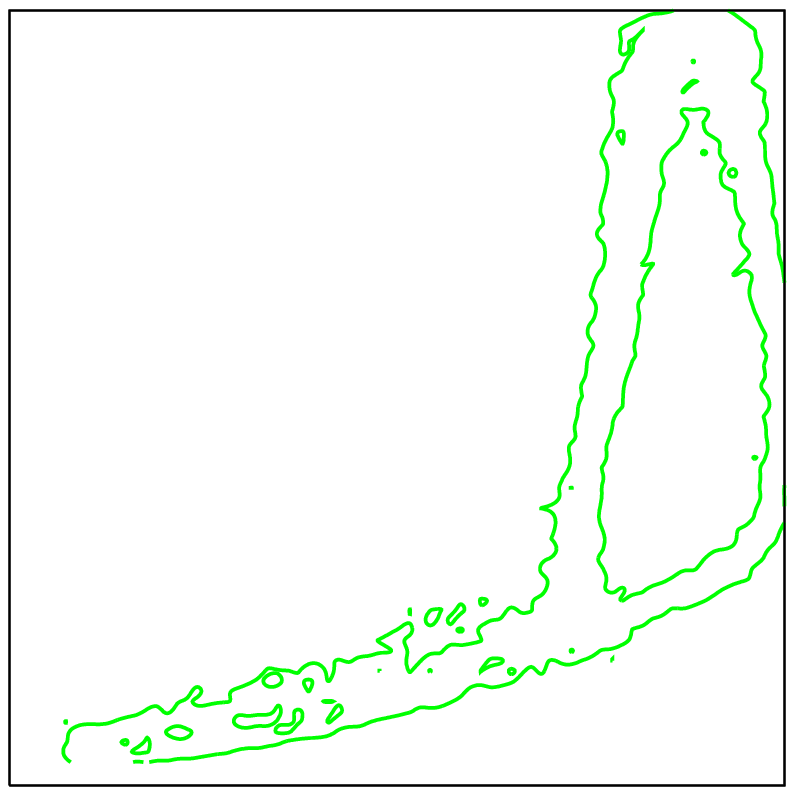}{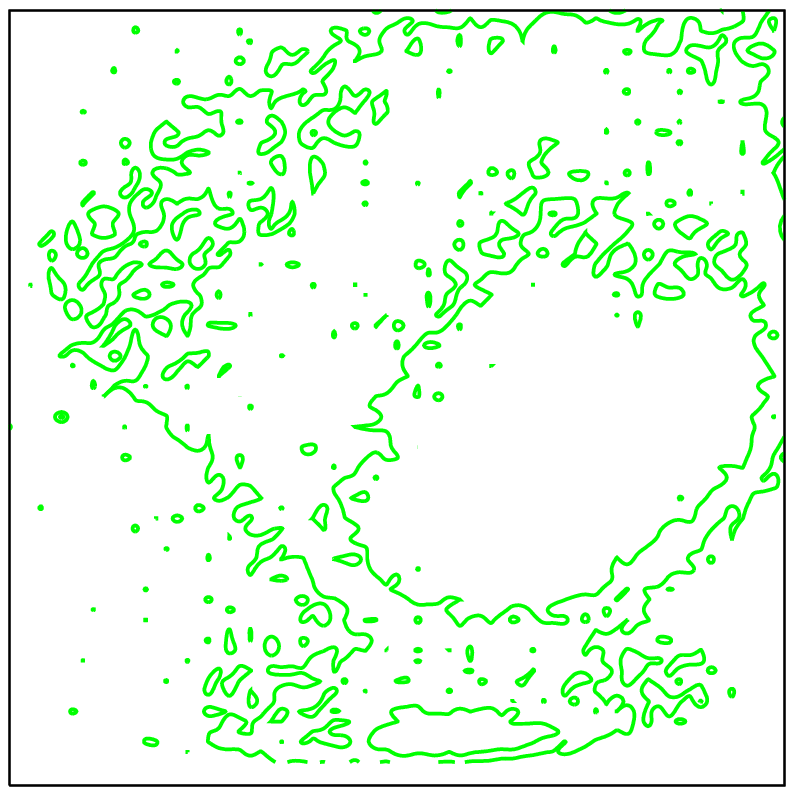}{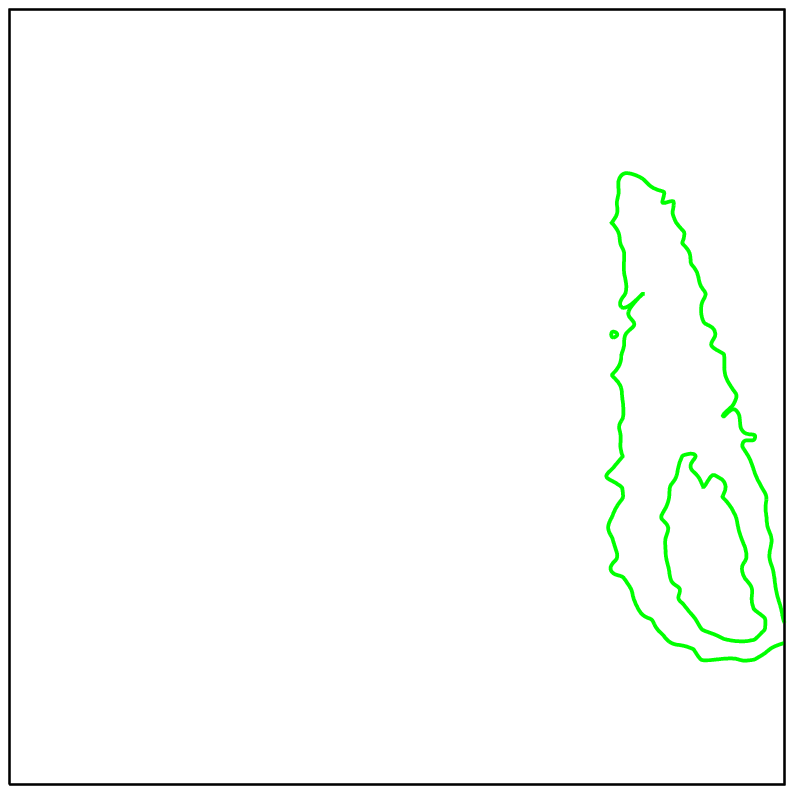}{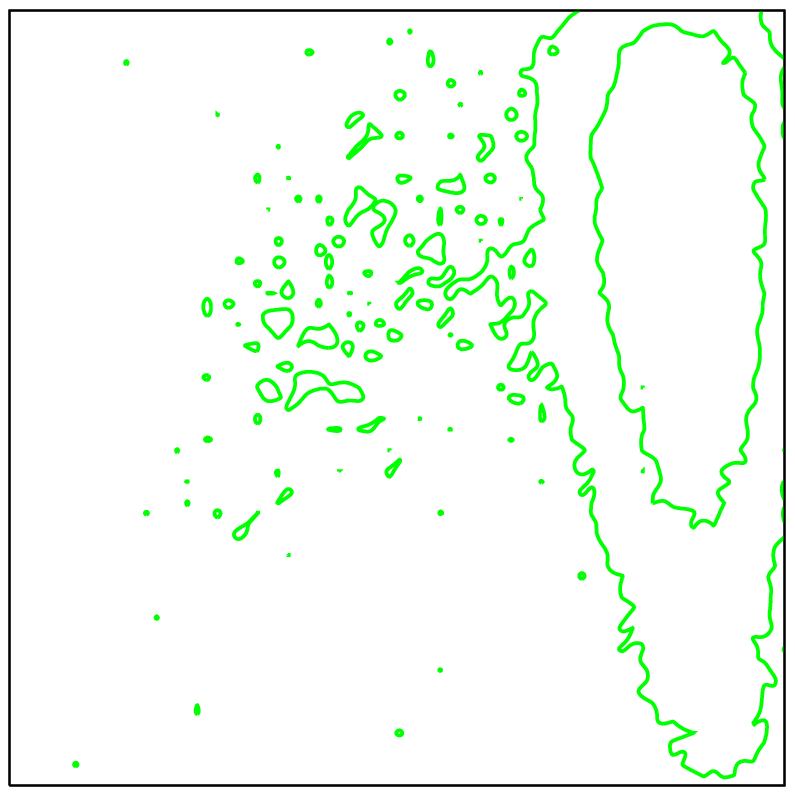}{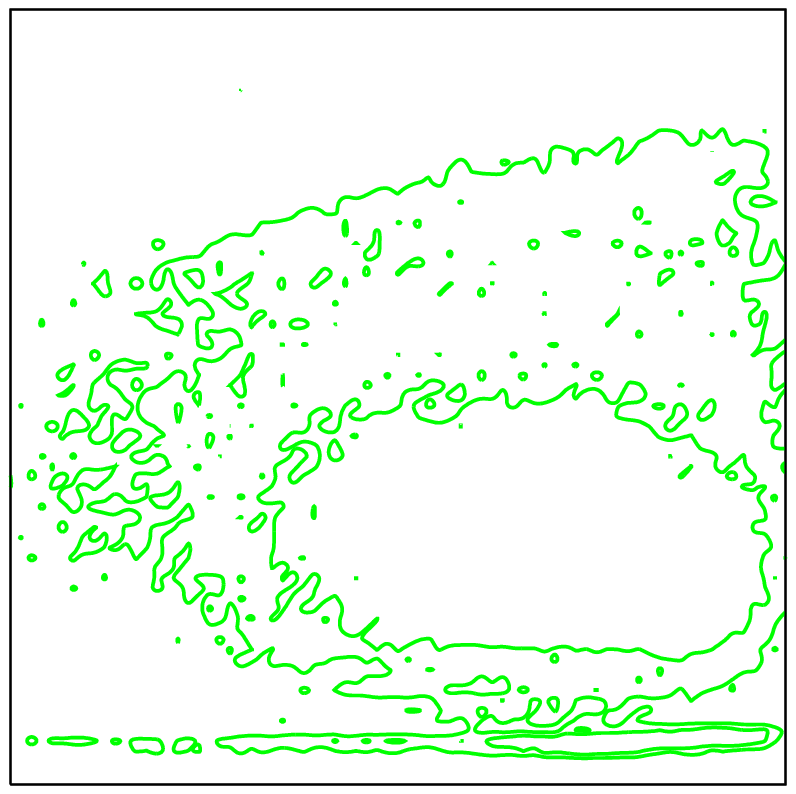}
\end{center}
\caption{Likelihood maps of mSUGRA parameter space.
The graphs show the
likelihood distributions sampled from
  7d parameter space and marginalised down to two.
The likelihood (relative to the likelihood in the highest bin) 
is displayed by reference to the bar on the right hand side of each plot.
The contours show the 68$\%$ and $95\%$ confidence level limits.
\label{fig:like7}}
\end{figure*}

We display binned sampled likelihood distributions in
Figs.~\ref{fig:like7}a-\ref{fig:like7}f for the full $m_t$, $m_b(m_b)$,
$\alpha_s(M_Z)^{\overline{MS}}$, $m_0$, $A_0$, $\tan \beta$ and $M_{1/2}$ parameter space. 
The unseen dimensions in each figure have been marginalised with flat priors. 
All 2-d or 1-d marginalisations in this paper assume a flat prior (within the
ranges of parameters considered in Table~\ref{tab:prior})
in all unseen dimensions (or in other words, the likelihood is integrated over
them with equal weight). One can view the marginalisation probabilistically,
or just as a means of viewing the higher dimensional parameter space. 
We have used 75$\times$75 bins, normalising the likelihood in each bin to the
maximum likelihood in any bin in each 2d plane.

In each plot, the $h^0$ pole $s$-channel resonant annihilation region
is present close to the lowest values of $M_{1/2}$. It can be seen as a
vertical sliver in the 
top-left hand corner of the $m_0-M_{1/2}$ as in Fig.~\ref{fig:like7}a
and the 
slim band ranging across the bottom of
Figs.~\ref{fig:like7}d,\ref{fig:like7}f.
The bright region in Fig.~\ref{fig:like7}a at low values of $m_0$ is
primarily a co-annihilation 
region where slepton-neutralino annihilation contributes significantly to the
depletion of the neutralino relic density in the early universe. However, at
the lowest values of $m_0$ and $M_{1/2}$ values visible on the graph, 
the bulk region resides, being continuously connected to the co-annihilation
tail, as shown in Ref.~\cite{Ellis:2003cw}.
The pseudoscalar Higgs ($A^0$) $s$-channel annihilation channel occurs at high
$\tan \beta=50-60$ and in the intermediate areas of $m_0=500-1600$ GeV,
$M_{1/2}=250-1400$ GeV. 
In the literature, the most common way to display mSUGRA results is to 
present them in 2d in the $m_0$-$M_{1/2}$ plane, where thin strips are
observed (see for example Ref.~\cite{Ellis:2004tc}) that are consistent with
the WMAP constraint upon $\Omega_{DM} h^2$. 
Fig.~\ref{fig:like7}a demonstrates (in corroboration with Refs.~\cite{Baltz:2004aw,Stark:2005mp})
that the strips are truly a result of
picking a 2d hyper-surface in parameter space: if one performs a full
multi-dimensional scan, there is a large region in the $m_0$-$M_{1/2}$ plane
that is 
consistent with the data. The bottom right hand side corner of
Fig.~\ref{fig:like7}a is ruled out primarily by the fact that the lightest
supersymmetric particle (LSP) is
charged. Large $M_{1/2}$ is disfavoured by the $(g-2)_\mu$ result. The bottom
left-hand corner of Fig.~\ref{fig:like7}a is ruled out by a combination of
dark matter and direct search constraints. We see an interesting correlation
between $m_0$ and $\tan \beta$ in Fig.~\ref{fig:like7}b: the region extending
to low $\tan \beta$ and low $m_0$ is essentially the stau co-annihilation/bulk
region.  

\begin{figure*}[t]
\begin{center}
\fourgraphs{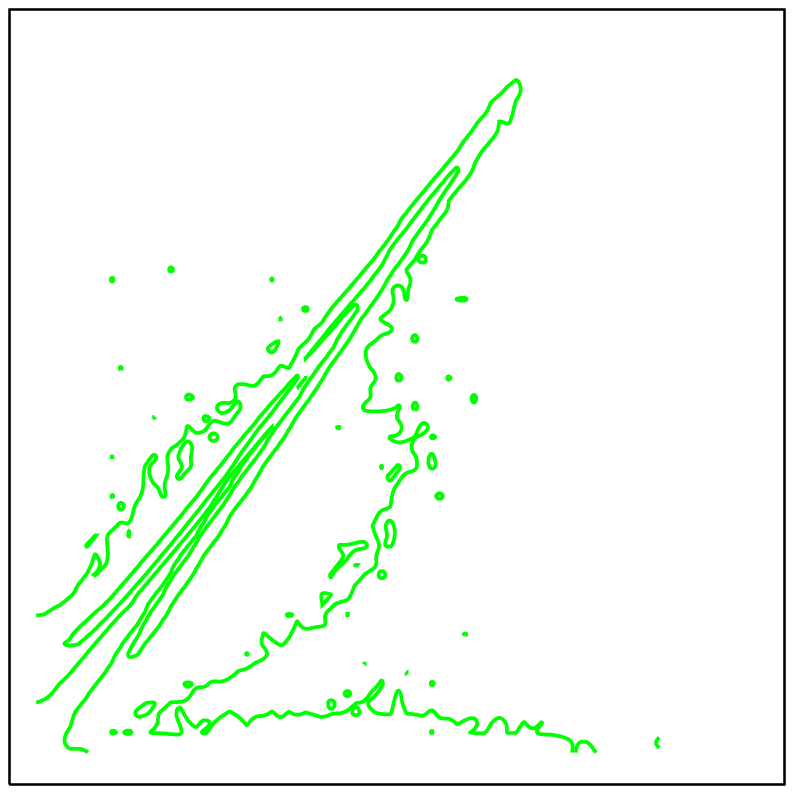}{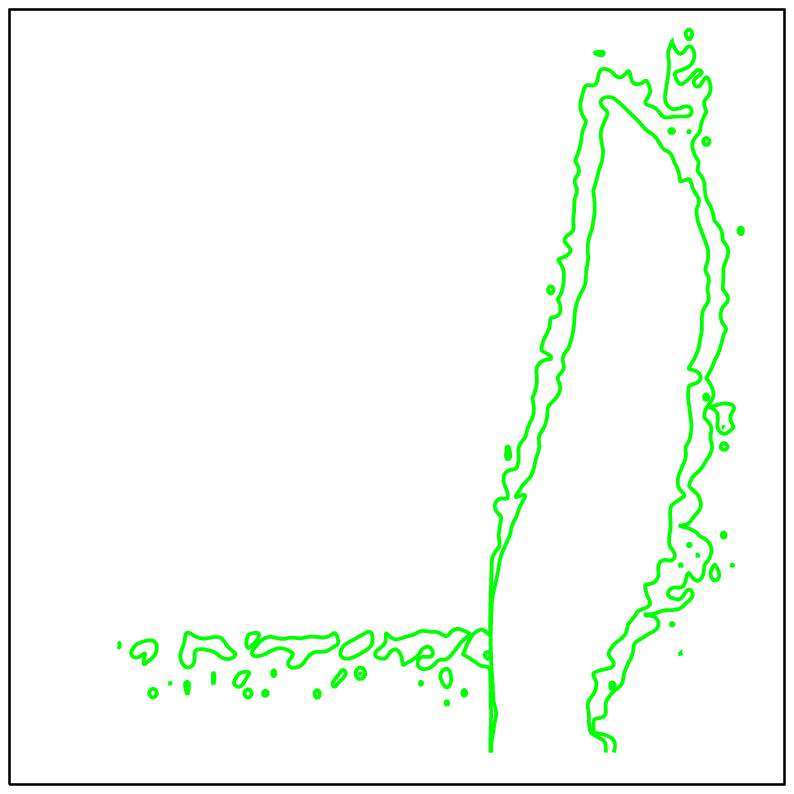}{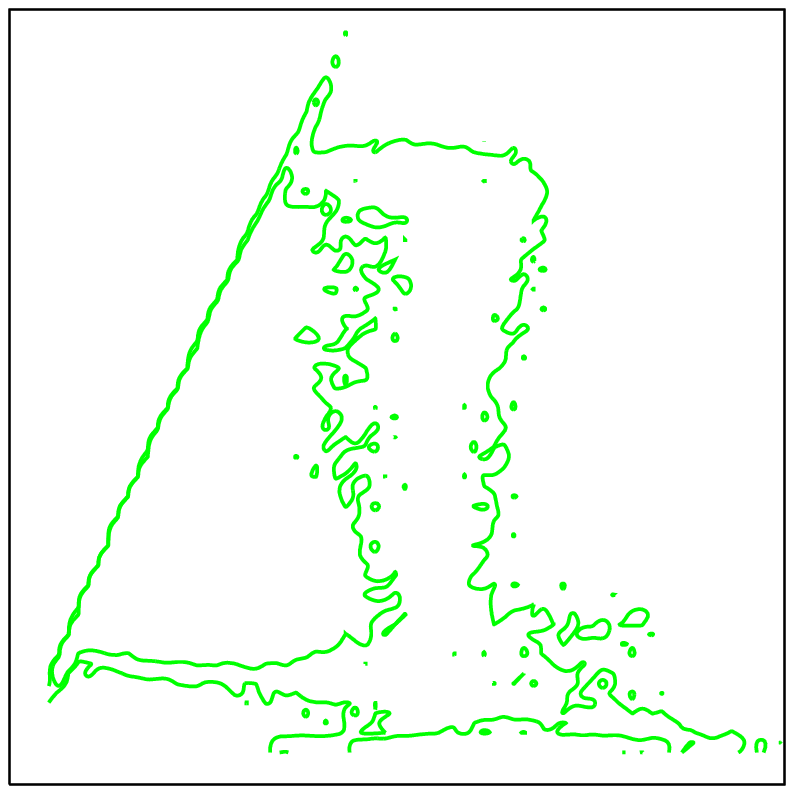}{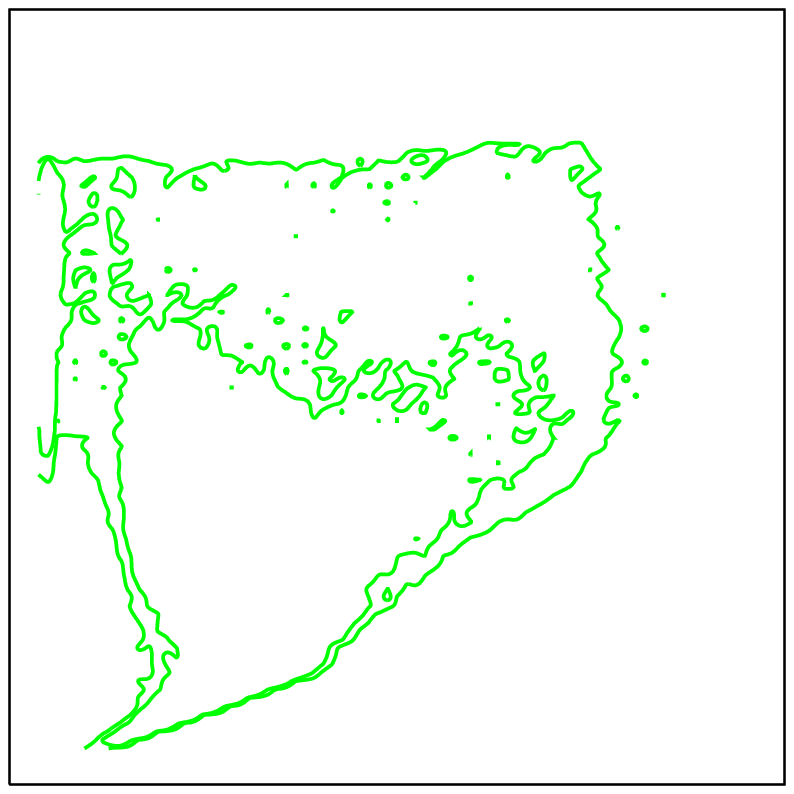}
\end{center}
\caption{Likelihood distributions of masses in mSUGRA\@. The graphs show the
likelihood distributions marginalised down to 2d.
The likelihood (relative to the likelihood in the highest bin) 
is displayed by reference to the bar on the right hand side of each plot.
The contours show the 68$\%$ and $95\%$ confidence level limits.
\label{fig:likeMass7}}
\end{figure*}
We display some binned likelihood distributions of MSSM particle masses
in
Figs.~\ref{fig:likeMass7}a-\ref{fig:likeMass7}c that are relevant for dark
matter annihilation. In 
Fig.~\ref{fig:likeMass7}a, the $A^0$-pole resonance region is clearly
discernable: at just above a line $2m_{\chi_1^0}=M_{A^0}$ and just below it,
there is 
just enough annihilation to produce the observed relic density. Throughout much
of the parameter space ($M_A<1$ TeV), the exact resonance condition depletes
the relic density
$\Omega_{DM} h^2$ to be too small. At around $M_A \sim 1$ TeV, exact resonance is
required in order to sufficiently deplete the relic dark matter density. 
The rest of the likelihood density is spread over the dark part of
the plot and is mostly too diffuse to be visible. 
In Fig.~\ref{fig:likeMass7}b, there are two detectable regions: the small one
with the maximum 2d binned likelihood at the lowest possible values
of $m_{\chi_1^0}$ corresponds to the $h^0$-pole region. The larger upper high
likelihood region
is an amalgam of the co-annihilation and $A^0$-pole regions.
As a by-product we see that values of $m_h>126$ GeV are disfavoured in the
mSUGRA model. 
In
Fig.~\ref{fig:likeMass7}c, the stau co-annihilation region is visible as the
diagonal dark line and the $h^0$-pole region as the lower likelihood horizontal
dark line. The rest of the likelihood density is diffusely distributed in
between these two extremes. 
We have shown the likelihood distribution for lightest chargino and slepton
masses in Fig.~\ref{fig:likeMass7}d. Unfortunately, most of the likelihood 
is in a region where the well-known tri-lepton search
channel~\cite{Abazov:2005ku} at the Tevatron 
is rather difficult. With 8 fb$^{-1}$ of integrated luminosity,
this search channel requires
$m_{\chi_1^\pm} < m_{\tilde l}$, $m_{\chi_1^\pm}<250$ 
for a discovery~\cite{future}. 

\begin{figure*}
\unitlength=1.1in
\begin{center}
\begin{picture}(5.8,6.0)(0,0.2)
\put(0,4.4){\epsfig{file=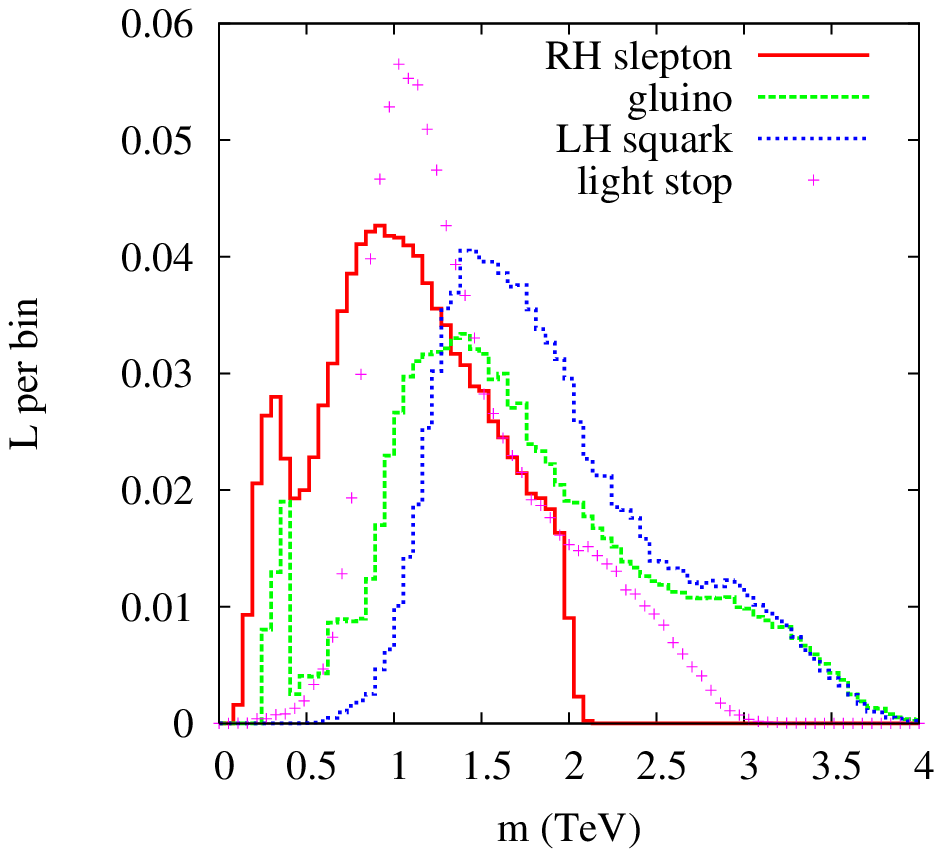, width=3.4in}}
\put(2.9,4.4){\epsfig{file=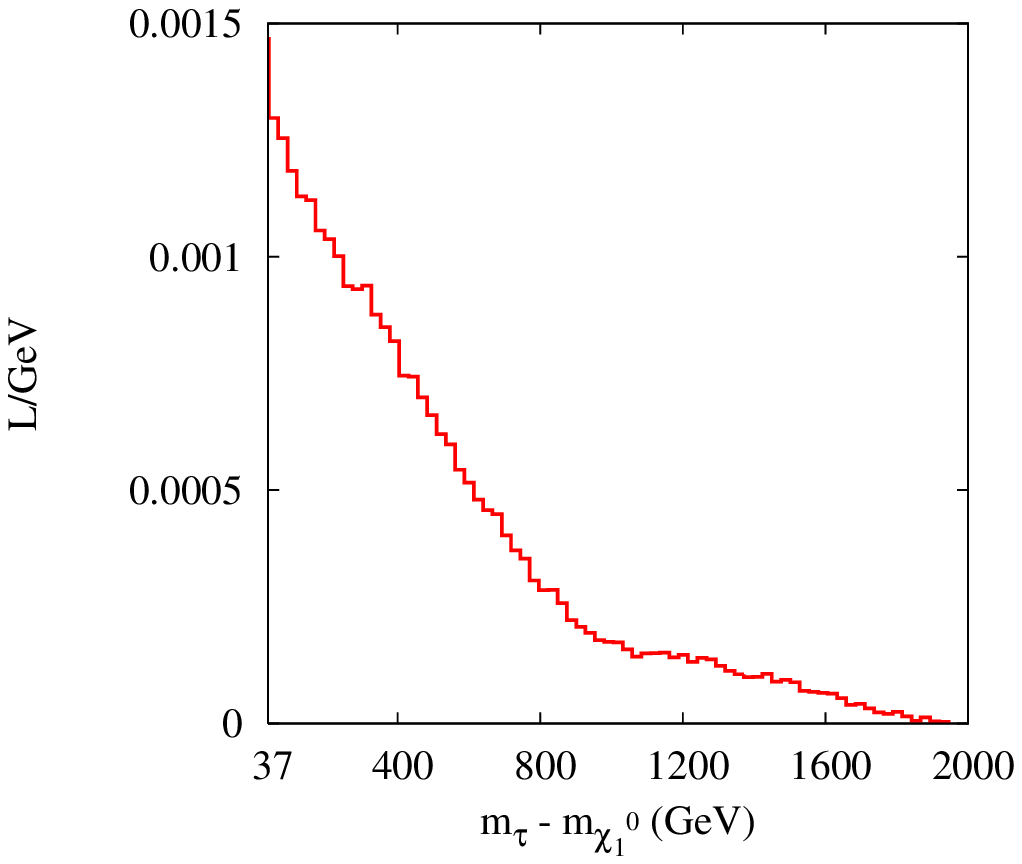, width=3.4in}}
\put(3.69,5.1){\epsfig{file=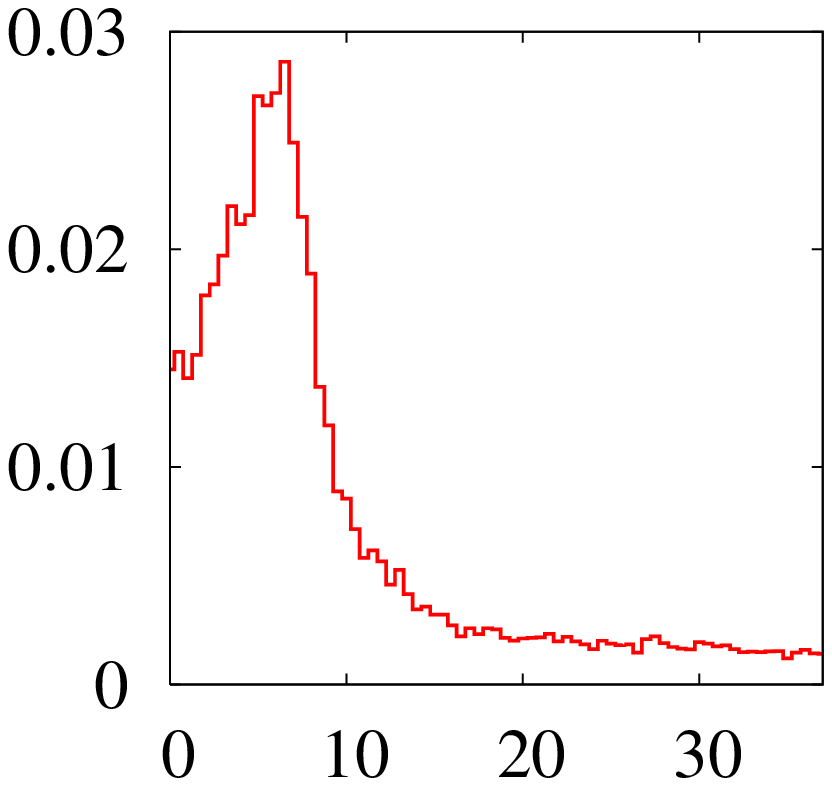, width=0.59\wth}}
\put(0,2.1){\epsfig{file=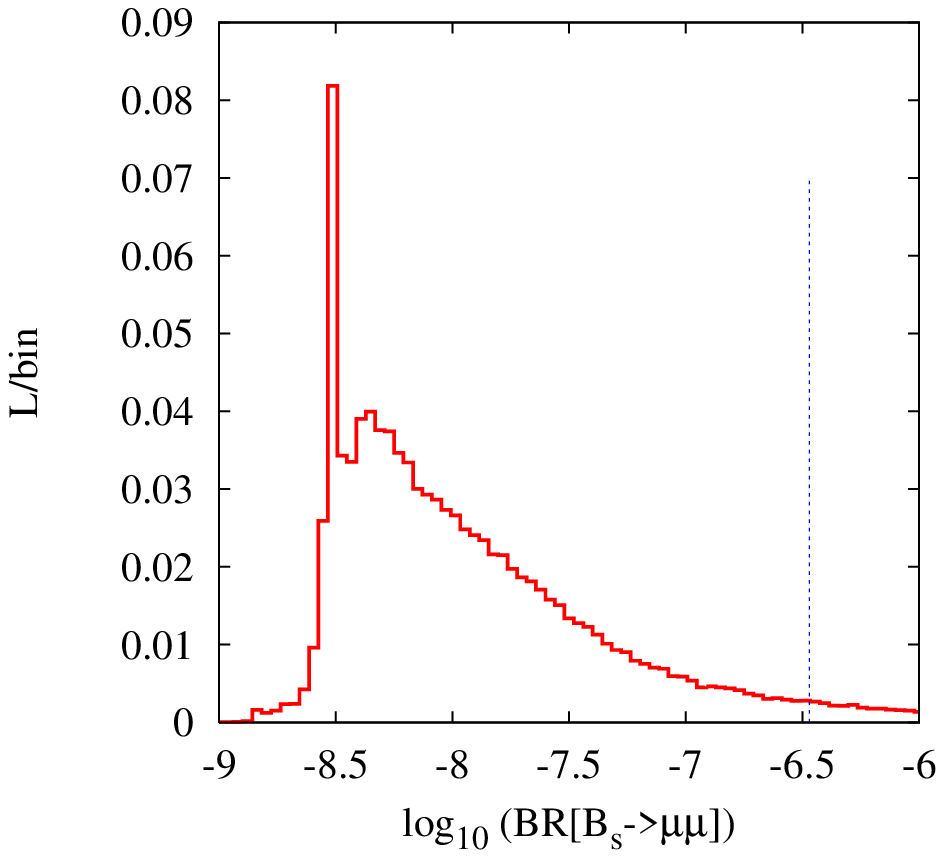, width=3.4in}}
\put(2.4,2.0){\epsfig{file=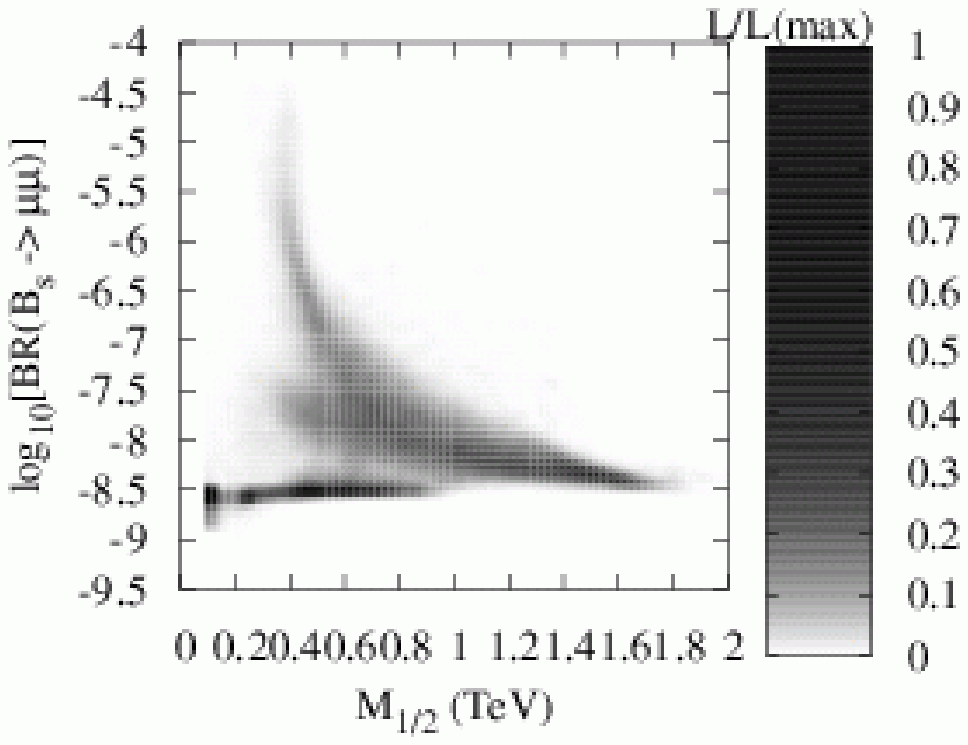, width=3.9in}}
\put(3.28,2.45){\epsfig{file=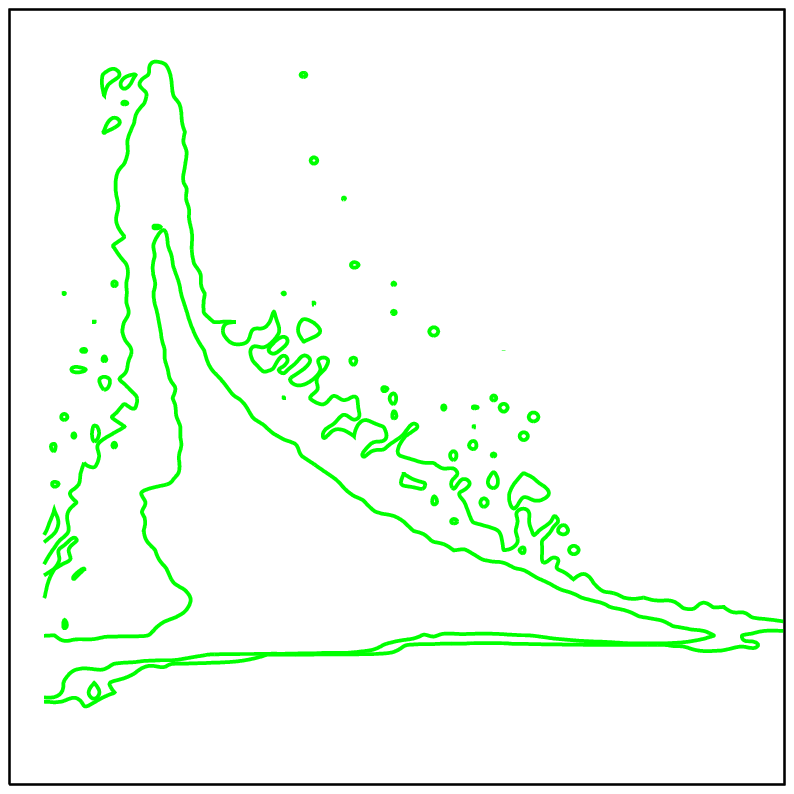, width=2.66in}}
\put(-0.5,-0.2){\epsfig{file=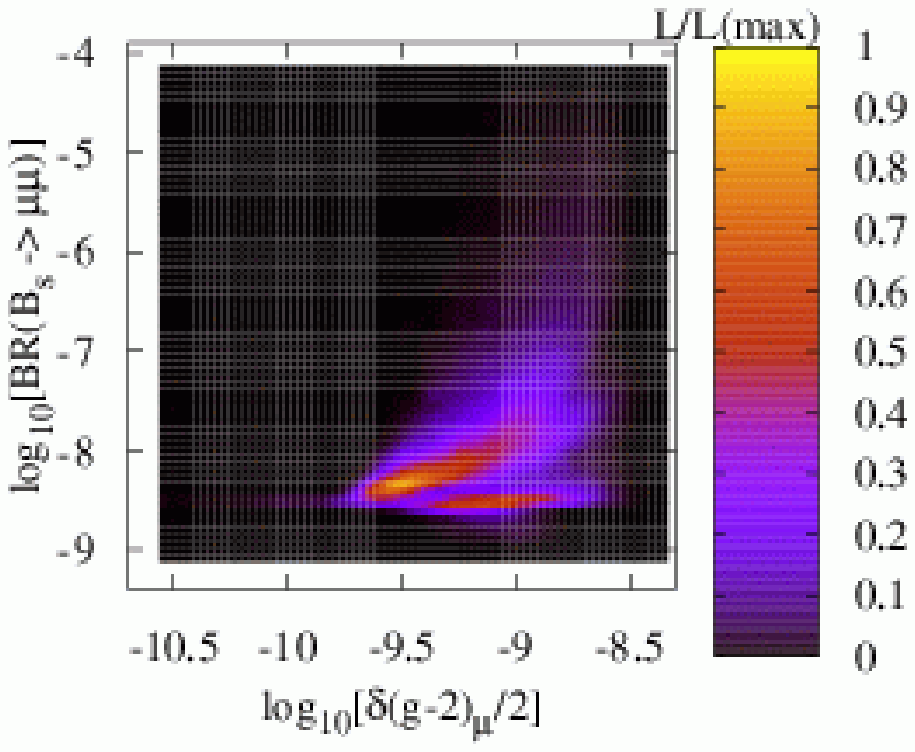, width=3.9in}}
\put(0.38,0.25){\epsfig{file=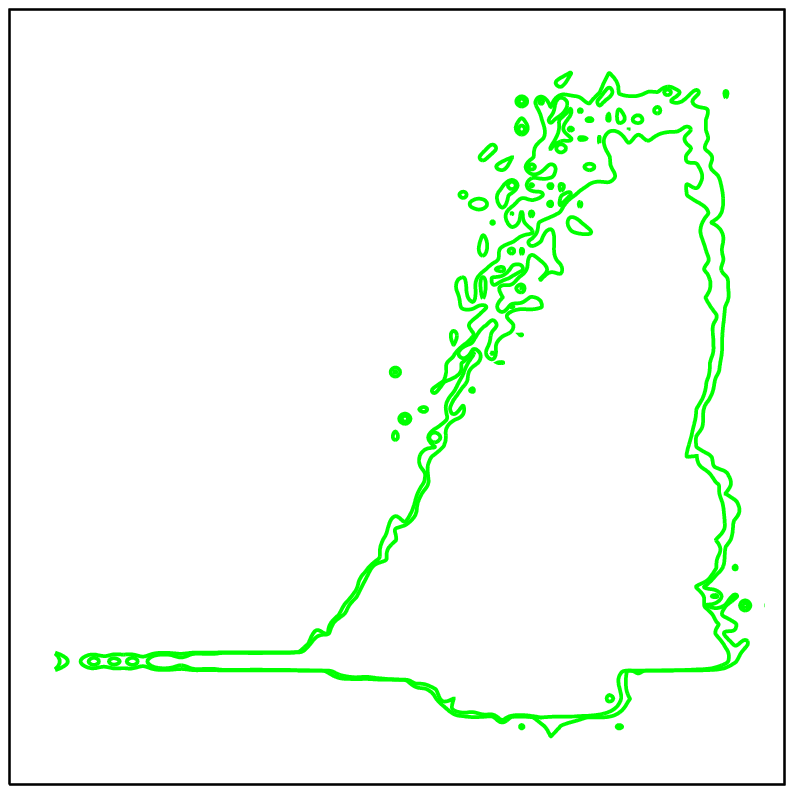, width=2.66in}}
\put(2.4,-0.2){\epsfig{file=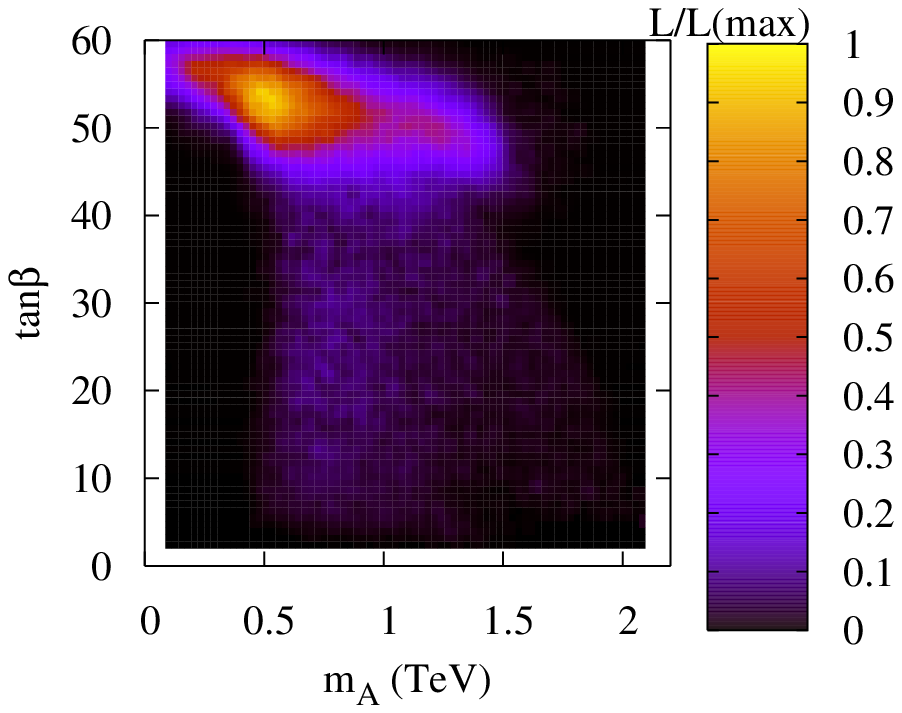, width=3.9in}}
\put(3.28,0.25){\epsfig{file=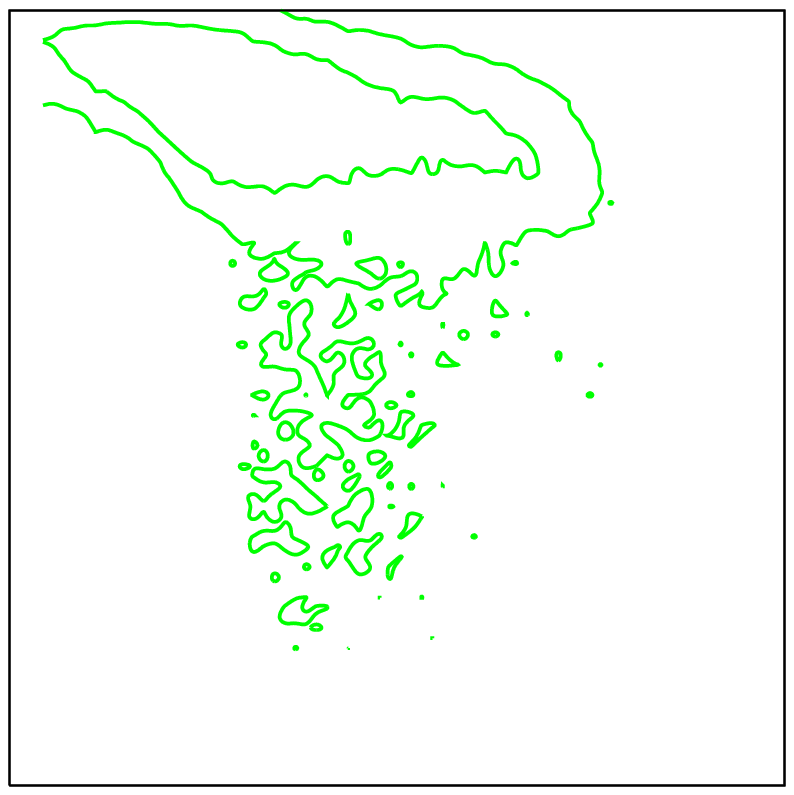, width=2.66in}}
\put(0,6.4){(a)}
\put(2.9,6.4){(b)}
\put(0,4.1){(c)}
\put(2.9,4.1){(d)}
\put(0,1.8){(e)}
\put(2.9,1.8){(f)}
\end{picture}
\end{center}
\caption{(a) Selected sparticle mass likelihood distributions in mSUGRA, 
(b) stau-neutralino mass difference likelihood distribution 
where the insert
  shows a blow-up of the quasi mass-degenerate region.
(c) branching ratio for the decay $B_s \rightarrow \mu^+ \mu^-$. 
The Tevatron upper bound is displayed by a vertical line.
(d) Likelihood density marginalised to the 2d plane
$BR(B_s \rightarrow \mu^+ \mu^-)$ versus $M_{1/2}$. (e)
Correlation between $BR(B_s \rightarrow \mu^+ \mu^-)$ and $(g-2)_\mu$.
(f) Likelihood marginalised to the $\tan \beta$-$M_A$ plane.
The contours show the 68$\%$ and $95\%$ confidence level limits.
\label{fig:hists}}
\end{figure*}

We show the sampled likelihood distribution for $m_{{\tilde l}_R}$,
$m_{\tilde g}$, $m_{{\tilde t}_1}$ and $m_{{\tilde q}_L}$ in
Fig.~\ref{fig:hists}a. The 
likelihood distributions have been placed into 75 bins of 
widths that are equal for the  
different types of sparticles. They are each normalised to an integrated
likelihood of 1. The spike at low values of the gluino mass corresponds to the 
$h^0$-pole region of mSUGRA\@. This spike has a good chance of being covered
at the Tevatron experiments before the LHC starts running~\cite{future}.
It should be noted that upper limits 
upon the scalar sparticle masses inferred from Fig.~\ref{fig:hists}a
are due largely to the definition of the range of
the initial parameters ($m_0$ being less than 2
TeV). Nevertheless, it is clear that there is already some preference from the
combined data for various ranges of sparticle masses and upper bounds upon the
gaugino masses. For example, values of $m_{\tilde g} > 3.5$ TeV and
$m_{\tilde g}=400-800$ GeV are disfavoured, as well as $m_{{\tilde
    q}_L}<800$ GeV. In Fig.~\ref{fig:hists}b, the mass splitting between the
lightest stau and lightest neutralino is shown. The insert in the figure
displays the quasi-degenerate co-annihilation region at low mass splittings. 
The peaked region at $m_{{\tilde \tau}_1} - m_{\chi_1^0}<$10 GeV is likely to
be difficult to discern at the Large Hadron Collider (LHC). 
One would wish to
measure decays of the 
lightest staus in order to check the co-annihilation region, but reconstructing
a relevant soft $\tau$ resulting from such a decay is likely to prove
problematic. On the other hand, a linear collider with
sufficient energy to produce sparticles could provide the necessary
information~\cite{Nojiri:1996fp,Bambade:2004tq}. 
The predicted likelihood distribution of the $B_s \rightarrow \mu^+ \mu^-$ branching
ratio is shown in Fig.~\ref{fig:hists}c. Possible values for this 
observable were 
found with a random scan of unconstrained MSSM parameter space in
Ref.~\cite{Dedes:2004yc} (no likelihood distribution was given).
The region to the right hand side of
the vertical line is ruled out from the combined
CDF/D0 95$\%$C.L. exclusion~\cite{Acosta:2004xj,Abazov:2004dj}\footnote{There
  are newer CDF/D0 bounds~\cite{Dagu}, for example CDF(D0) have non-combined
  95$\%$ C.L. limits of 2.0(3.0)$\times 10^{-7}$ respectively.}
limit $BR(B_s^0 \rightarrow \mu^+
\mu^-) < 3.4 \times 10^{-7}$. 
We have not cut points violating this
constraint, but this has negligible effect
since there are
only a small number of them. The 2d marginalisation of
the branching ratio versus $M_{1/2}$ shows a peak at very low $M_{1/2}$
values, as Fig.~\ref{fig:hists}d displays.
This indicates that the spike in Fig.~\ref{fig:hists}c at branching ratios
of about $10^{-8.6}$ is due to the $h^0$-pole region. 
Lowering the empirical
upper bound on the $B_s \rightarrow \mu^+ \mu^-$ branching ratio will
significantly cut into the allowed mSUGRA parameter space. 
A significant lowering of the bound upon this branching ratio is expected 
in the coming years from the Tevatron experiments and from LHCb.
For example, it is estimated~\cite{future} that the Tevatron could exclude 
a branching ratio of more than $2 \times 10^{-8}$ with 8 fb$^{-1}$ of
integrated luminosity. This corresponds to ruling out 29$\%$ of the
currently allowed likelihood density. 
Predictions for $BR(B_s \rightarrow \mu^+ \mu^-)$ were correlated
with those for $(g-2)_\mu$ in Ref.~\cite{Dedes:2001fv} in mSUGRA\@. For a given
mSUGRA 
parameter point, a correlation was shown when $\tan \beta$ was varied. 
The authors conclude that for high $\tan \beta$, 
an enhancement of 
$BR(B_s \rightarrow \mu^+ \mu^-)$
is implied by the $(g-2)_\mu$ measurements. We re-examine this statement in view
of the full dimensionality of the mSUGRA parameter space in
Fig.~\ref{fig:hists}e. The correlation is seen to be far from perfect, the
likelihood distribution being highly smeared in terms of the two measurements.
Nevertheless, there is a mild correlation between $BR(B_s \rightarrow \mu^+
\mu^-)$ and the SUSY contribution to $(g-2)_\mu$ 
at the highest likelihoods, as evidenced by the bright
oblique stripe in Fig.~\ref{fig:hists}e.
In Fig.~\ref{fig:hists}f, we show the likelihood distribution in the $\tan
\beta$-$m_A$ plane. There is a significant amount of likelihood density
towards the top left-hand side of the plot, where the Tevatron is expected to
have sensitivity~\cite{future} (covering $\tan \beta>40$ for $m_A<240$ GeV for
8fb$^{-1}$ of integrated luminosity). LHC experiments should be able to 
observe the $A^0$ for
the entire high likelihood range~\cite{Armstrong:1994it}.

We now ask what are the likelihoods of the 
different regions of relic density depletion in mSUGRA parameter
space. In order to sharply delineate the
regions, 
we define them as follows: 
the co-annihilation region is defined such that $m_{\chi_1^0}$ is within $10\%$
of $m_{{\tilde \tau}_1}$.
The $h^0/A^0$-pole regions have $2m_{\chi_1^0}$ within $10\%$ of $m_{h^0}$,
$m_{A^0}$ respectively.
Stop co-annihilation requires a broader definition: it is defined such that 
$m_{\chi_1^0}$ is within 30$\%$ of $m_{{\tilde t}_1}$, since the
annihilation is so much
more efficient~\cite{Boehm:9911,arnie,Ellis:2001nx} than in the other regions.
A better defined procedure might perhaps be to determine regions on the basis
of the {\em dominant}\/ annihilation mechanism, but since we are only looking
for a rough indication of the region involved, the procedure adopted here
will suffice. 
Points that fall in between any of the sharp definitions are either 
from the bulk region or in the smaller tails of the likelihood distribution.

\begin{table}
\caption{Likelihood of being in a certain region of mSUGRA parameter
  space. \label{tab:regions}}
\begin{tabular}{|c|c|}
\hline Region & likelihood \\ \hline
$h^0$ pole & 0.02$\pm$0.01 \\
$A^0$ pole & 0.41$\pm$0.03 \\
$\tilde \tau$
co-annihilation & 0.27$\pm$0.04 \\ 
$\tilde t$ co-annihilation & $(2.1\pm4.8)\times 10^{-4}$\\ \hline
\end{tabular}
\end{table}
The likelihoods of these regions are shown in Table~\ref{tab:regions}.
We estimate the uncertainty by calculating the standard deviation on the 9
independent Markov chain samples. The quoted error thus reflects an
uncertainty due not to experimental errors, but to a to finite
simulation time of the Markov chain.
We see that the $h^0$-pole region has a relatively low likelihood whereas for
the $A^0$-pole and $\tau$ co-annihilation regions the likelihood is larger. 
From the table, we see that the $\tilde t$-co-annihilation region, although
uncertain due to the low statistics, is
negligible, and we now investigate why this is the case.

\begin{figure*}
\unitlength=1.1in
 \begin{picture}(5.8,2.6)(0.5,0)
 \put(0.2,0){\epsfig{file=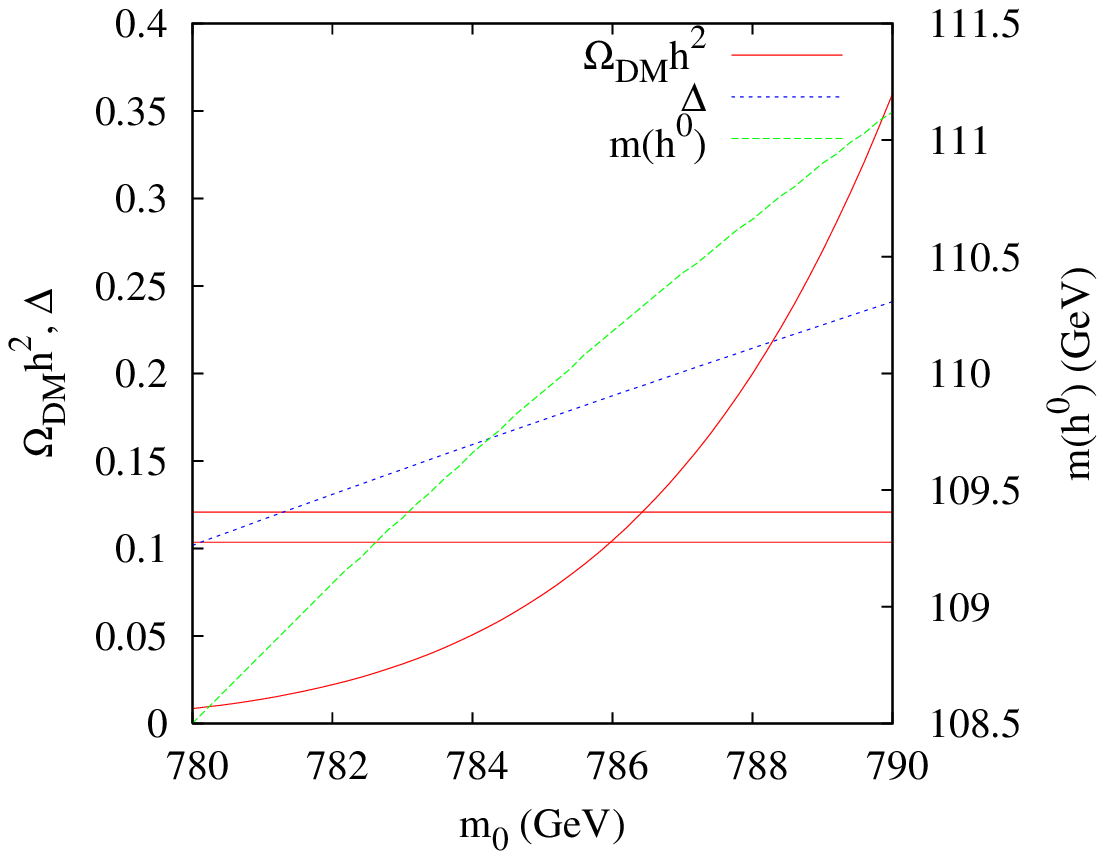, width=0.9\wth}}
 \put(3.2,0){\epsfig{file=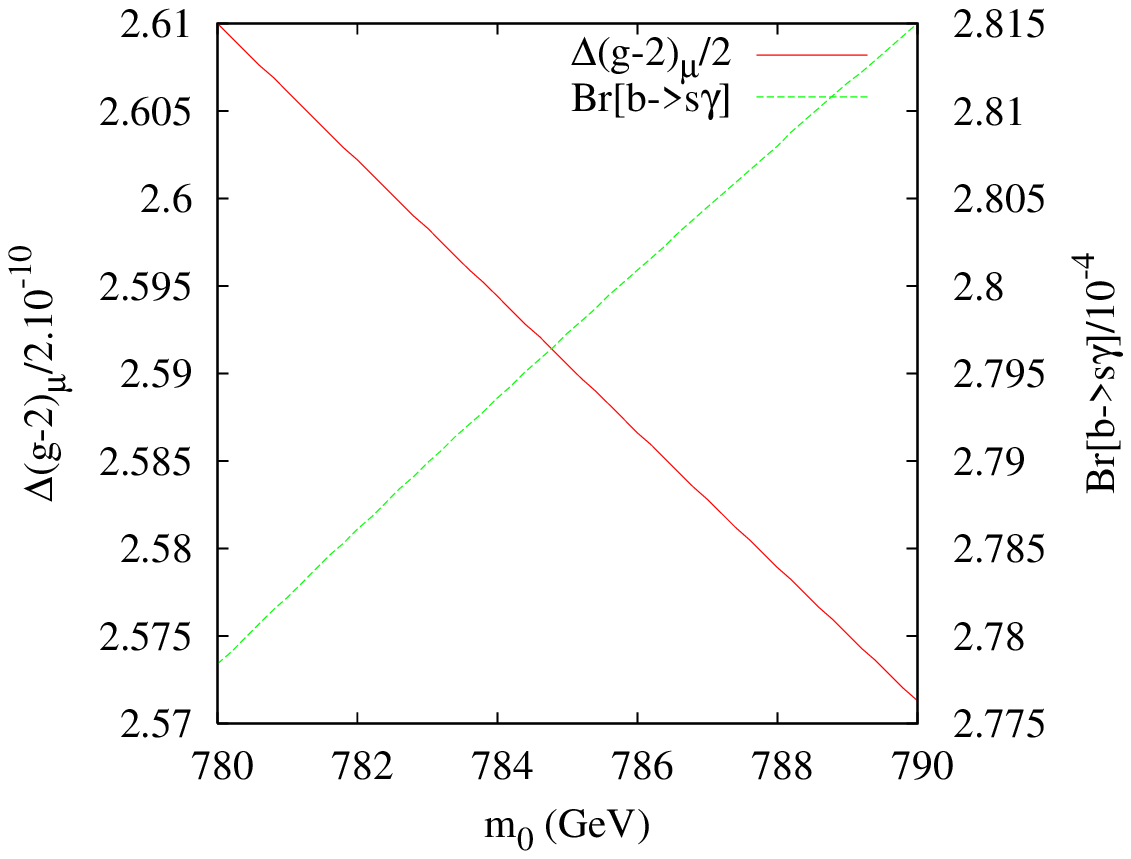, width=0.9\wth}}
 \put(0.2,2.1){(a)}
 \put(3.1,2.1){(b)}
 \end{picture}
\caption{A line through the stop co-annihilation region in mSUGRA: $\tan
  \beta=10$,  $M_{1/2}=350$ GeV, $A_0=-2000$ GeV and central Standard Model
  inputs. (a) dark matter relic density, fractional stop-neutralino mass
  splitting $\Delta \equiv (m_{{\tilde t}_1} - m_{\chi_1^0}) / m_{\chi_1^0}$ 
  and light Higgs mass. The horizontal lines show the 1$\sigma$ WMAP-derived
  limits upon $\Omega_{DM} h^2$ derived from Eq.~\protect\ref{omConst}, (b)
  muon magnetic moment and $BR[b \rightarrow s \gamma]$ predictions. 
\label{fig:stopCoan}}
\end{figure*}

The suppression of the stop-co-annihilation region comes from essentially two
effects: firstly, as already apparent from Ref.~\cite{Ellis:2001nx}, finding a
suitable stop co-annihilation region which is compatible with {\em both} the
$(g-2)_\mu$ and $BR[b   \rightarrow s \gamma]$ measurements is problematic. 
Secondly, the central value of $m_t$ has come down since
ref.~\cite{Ellis:2001nx}. The dominant radiative corrections to $m_{h^0}$ are 
highly correlated with $m_t$~\cite{Allanach:2004rh}, 
with the consequence that the lower predicted Higgs mass 
now rules out more of the stop co-annihilation region. 
We illustrate these points in Fig.~\ref{fig:stopCoan} along the $m_0$
direction for given values of the other mSUGRA parameters (stated in the
caption). 
In Fig.~\ref{fig:stopCoan}a, we plot the fractional stop-neutralino
mass splitting $\Delta$ 
alongside the neutralino relic density $\Omega_{DM} h^2$. 
We see that the fractional mass splitting takes values between 0.1 and 0.23 in
the range of $m_0$ shown. This is the stop-co-annihilation regime, and we see
that around $m_0 \sim 786$ GeV, $\Delta \sim 0.2$ corresponding to
$\Omega_{DM} h^2$ roughly compatible with the WMAP constraint. Unfortunately,
we also see that the lightest CP-even Higgs mass is predicted to be 110.3
GeV here and is ruled out by the LEP2 Higgs limits shown in Table~\ref{tab:par}
($\sin^2 (\beta-\alpha) =1.0$) for this range of parameters). This problem is
remedied by going to higher values of $m_t$, since $m_{h^0}$ then goes up, but
of course this comes with an associated penalty in the likelihood from being
away from the empirically central values of $m_t$. 
In Fig.~\ref{fig:stopCoan}b, we display predictions for $BR[b\rightarrow s
  \gamma]$ and $\delta (g-2)_\mu/2$ along the chosen range for $m_0$. 
These predictions are both lower than the empirically derived constraints in
the region where the dark matter relic density is in accordance with the WMAP
constraint: including errors in the theoretical prediction as described in the
previous section, 
$BR[b \rightarrow s \gamma]$ is 1.95$\sigma$ lower than the
central value and $\delta (g-2)_\mu/2$ is 1.6$\sigma$ lower. It turns out that
these predictions
are {\em not} very sensitive to changes in $m_t$ and so their likelihood
penalty tends to apply for the higher values.
However, lower values of $(-A_0)$
require lower $m_0$ in order to fit the dark matter constraint and pick up 
a bigger likelihood constraint from the egregious prediction of
$BR[b\rightarrow s \gamma]$.  
These findings are in rough agreement with those of Ref.~\cite{Ellis:2001nx}
except for the more restrictive Higgs bounds, which are a consequence of the
lower experimental value of $m_t$. Ref.~\cite{Ellis:2001nx} only applies
2$\sigma$ bounds on both $(g-2)_\mu$ and $BR[b \rightarrow s \gamma]$, whereas
our results take into account the likelihood penalty paid by the fact that
neither prediction is close to its central value near the stop co-annihilation
region, which then becomes disfavoured compared to the other regions (where
an almost perfect fit is possible). 
There is also a volume effect: in Table~\ref{tab:regions}, the likelihoods we 
calculate are {\em integrated} over the relevant region. Thus regions that are
very small, such as the stop co-annihilation region, will tend to have a smaller
likelihood than other, larger regions. 
In analyses in following sections, stop 
co-annihilation also turns out to have negligible likelihood and so we will 
neglect it from the results. 

As an aside, we note that the decay chain ${{\tilde q}_L} \rightarrow
{\chi_2^0} \rightarrow {\tilde l}_R \rightarrow \chi_1^0$ exists with a
likelihood  
of $0.24\pm0.04$ (this number is just based upon the mass ordering and does
not take into account the branching ratio for the chain). The existence of
such a chain 
allows the extraction of several functions of sparticle masses from kinematic
end-points and they have been used in many LHC analyses,
for example Refs.~\cite{Armstrong:1994it,Allanach:2000kt,Weiglein:2004hn}.

\section{Theoretical Uncertainty \label{sec:uncertainties}}

\begin{table}
\caption{Likelihood of being in a certain region of mSUGRA parameter
  space including theoretical uncertainties in the sparticle spectrum
  calculation. \label{tab:regions2}} 
\begin{tabular}{|c|c|}
  \hline Region & likelihood \\ \hline
$h^0$ pole & 0.03$\pm$0.01 \\
$A^0$ pole & 0.41$\pm$0.05 \\
co-annihilation & 0.26$\pm$0.08 \\ \hline
\end{tabular}
\end{table}
Theoretical uncertainties in the sparticle mass predictions have been shown to
produce non-negligible effects in fits to data~\cite{Allanach:2003jw},
including fits to the relic
density~\cite{Allanach:2004jh,Allanach:2004xn,Belanger:2005jk}. 
In this section, we perform a second MCMC analysis taking theoretical
uncertainty into account in order to estimate the size of its effect.
{\tt SOFTSUSY} performs the Higgs potential minimisation then 
calculates sparticle pole masses at a scale $M_{SUSY}=\sqrt{m_{{\tilde t}_1}
  m_{{\tilde t}_2}}$. This scale is chosen because it is hoped that loop
corrections to the pole mass corrections that are not yet taken into account
(typically two-loop corrections) are small at this scale. 
In order to estimate the size of theoretical uncertainties, we vary this
scale by a factor of two in either direction (but it is always constrained 
to be greater than $M_Z$)\footnote{A more complete estimate would be to 
 calculate the Higgs
potential minimisation conditions and the sparticle
masses all at different scales, varying each independently by a factor of
two. However, such a prescription is impractical here due to CPU time
constraints.}. 
Implementation of the uncertainty in the MCMC algorithm is
simple: we simply add an input
parameter $x$ which is bounded between 0.5 and 2, giving the factor by which
$M_{SUSY}$ is to be multiplied. The MCMC is then re-run as before and explores
the  full 8d parameter space (including $x$) accordingly. 

Such a
procedure automatically takes into account the correlations in predictions due
to correlated theoretical uncertainties in the sparticle mass predictions.
The likelihoods of the three mSUGRA regions are shown in
Table~\ref{tab:regions2}. The likelihoods are approximately equal to those for
the 7d case, as a comparison to Table~\ref{tab:regions} indicates.
In fact, comparing
results and plots produced with and without theoretical uncertainties, 
the results are generally very similar. 
This indicates that the theoretical uncertainties don't
make a huge difference to the 1d and 2d marginalisations
compared to those coming from the data.
The decay chain ${{\tilde q}_L} \rightarrow
{\chi_2^0} \rightarrow {\tilde l}_R \rightarrow {\chi_1^0}$ has a likelihood
of $0.22 \pm 0.08$, not significantly 
different to the case when theoretical uncertainties were not taken into
account. 

\begin{figure*}
\twographst{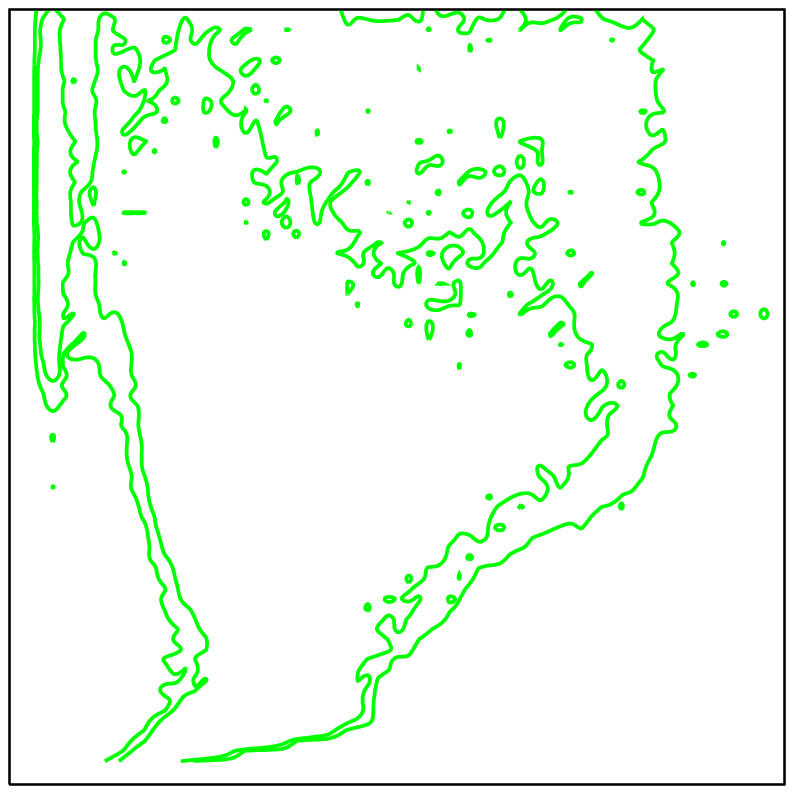}{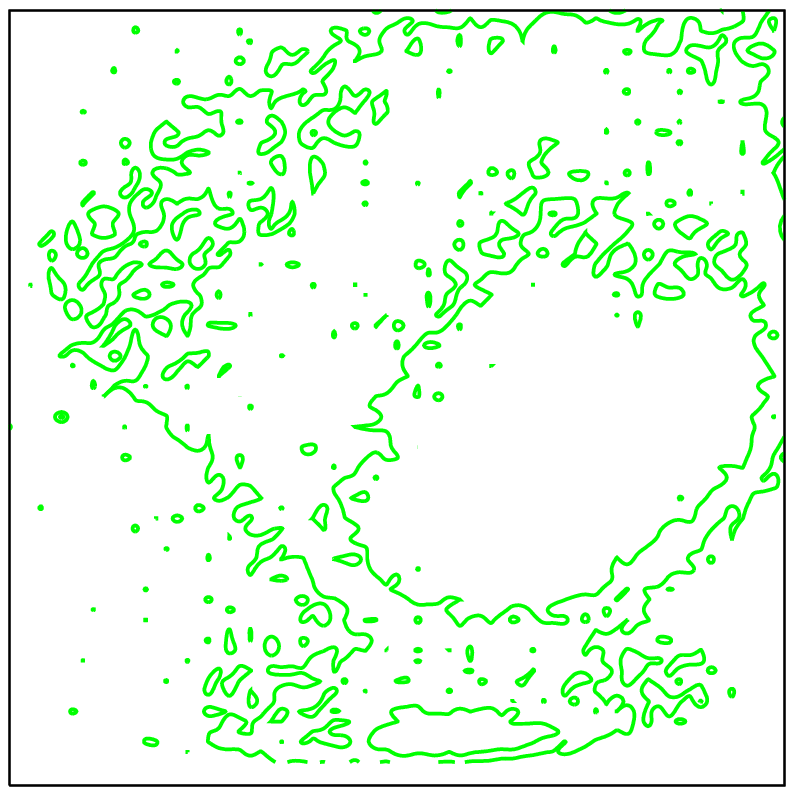}
\caption{Likelihood maps of mSUGRA parameter space including theoretical
  uncertainty. 
The graphs show the
likelihood distributions sampled from
  8d parameter space and marginalised down to two.
The likelihood (relative to the likelihood in the highest bin) 
is displayed by reference to the bar on the right hand side of each plot.
The contours show the 68$\%$ and $95\%$ confidence level limits.
\label{fig:uncer}}
\end{figure*}

We show
two of 
the  2d marginalised likelihoods in
Figs.~\ref{fig:uncer}a,\ref{fig:uncer}b. 
Comparing with Fig.~\ref{fig:like7}, we see that Fig.~\ref{fig:uncer} shows no
significant effects deriving from theoretical errors.
Marginalising mass likelihood distributions down to 1d (as in
Fig.~\ref{fig:hists}a for example), one obtains distributions that are
identical to  the 7d case except for small statistical fluctuations. 

\section{Other Sources of Dark Matter\label{sec:other}}
\begin{figure*}
\twographs{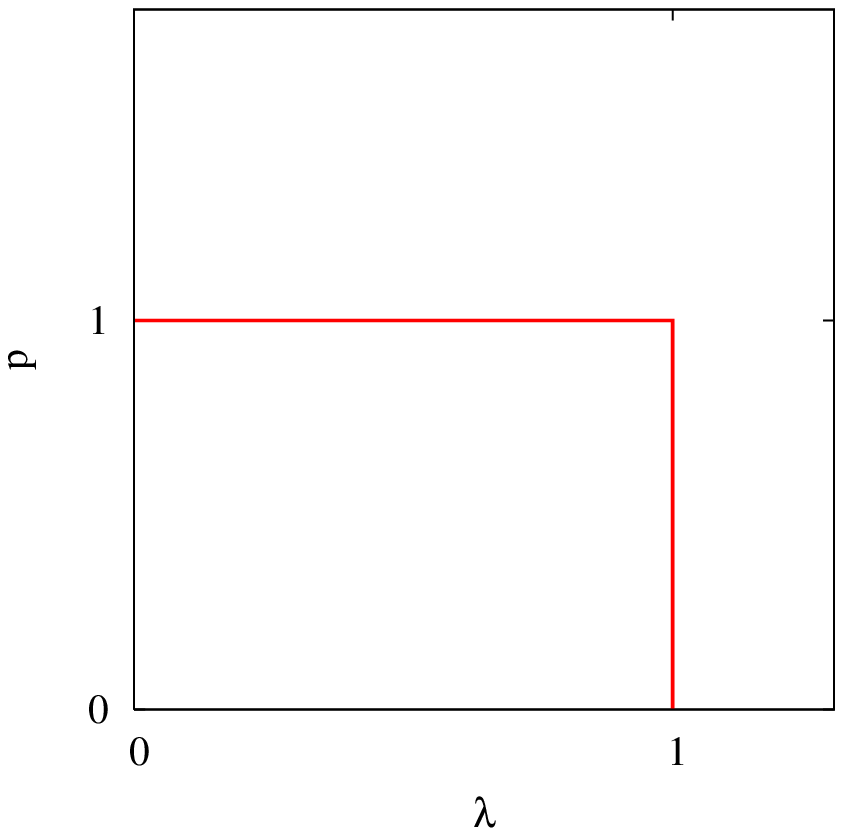}{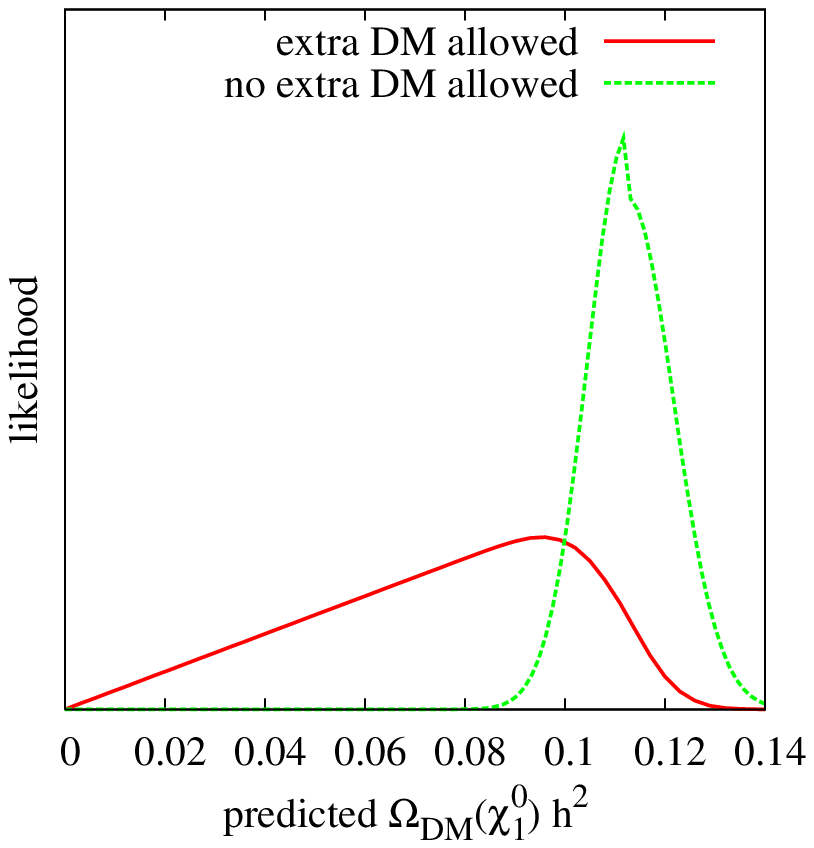}
\caption{(a) $p(\lambda | m_{\Omega_{DM} h^2})$, the probability distribution
  assumed for $\lambda$,  
the ratio of predicted SUSY
dark matter to total dark matter, 
given some measured value $m_{\Omega_{DM} h^2}$. 
(b) Comparison of the likelihood     penalty 
  ${\mathcal L}_{\Omega_{DM} h^2}$
    paid for a prediction of SUSY dark matter $p_{\Omega_{DM} h^2}$ with 
(``extra DM allowed'') and
      without (``no extra DM allowed'') 
      the allowance of non thermal-neutralino dark matter. \label
    {fig:prior}}
\end{figure*}

For completeness, we may ask how robust our results are with respect to the
assumption that non thermal-neutralino contributions to $\Omega_{DM}$ are
negligible.  
Thus, we allow the predicted amount of thermal neutralino relic density
$p_{\Omega_{DM} h^2}$ (assuming some mSUGRA point $s$) to be some fraction of
the total relic predicted relic density $\Omega_{DM}^{tot} h^2$:
\begin{equation}
\lambda = \frac{p_{\Omega_{DM} h^2}}{\Omega_{DM}^{tot} h^2}, \qquad 0 \leq
\lambda \leq 1.
\end{equation}
Thus, the total amount of dark matter predicted is 
$p_{\Omega_{DM} h^2} / \lambda$ and,
assuming a flat pdf for $\lambda$ as shown in Fig.~\ref{fig:prior}a,
we obtain the likelihood penalty for a given SUSY dark matter prediction
\begin{widetext}
\begin{equation}
{\mathcal L}_{\Omega_{DM} h^2} = 
\int^1_0d \lambda \ 
\frac{1}{\sqrt{2 \pi} s_{\Omega_{DM} h^2}(\lambda)}
\exp \left[ \frac{-(m_{\Omega_{DM} h^2} - 
p_{\Omega_{DM} h^2} / \lambda)^2}{2 s_{\Omega_{DM} h^2}(\lambda)^2}
\right], \label{lowOm}
\end{equation}
\end{widetext}
where 
\begin{eqnarray}
s_{\Omega_{DM} h^2}(\lambda) &\equiv& 
0.0081\ \theta(p_{\Omega_{DM} h^2} /
\lambda - m_{\Omega_{DM} h^2}) +\nonumber \\
&&
0.0091\ \theta(-p_{\Omega_{DM} h^2} /
\lambda + m_{\Omega_{DM} h^2})
\end{eqnarray}
in accordance with the asymmetric errors in Eq.~\ref{omConst}.
$\theta(x)$ is the Heavisde step function, $\theta(x)=1$ for all $x\geq 0$,
$\theta(x)=0$ for all $x<0$.
We calculate Eq.~\ref{lowOm} numerically and display it in
Fig.~\ref{fig:prior}b, where it is contrasted with the old dark matter
likelihood penalty that assumes that all dark matter is of thermal neutralino
origin. 
The figure shows that if the relic density is too {\em high}, a severe
likelihood penalty applies (similar to the ``no extra DM allowed'' case)
but a much less severe penalty applies
if the prediction is below the central value
of the observed WMAP value in Eq.~\ref{omConst}. 
The additional contribution to $\Omega h^2$ is assumed 
to be provided by some non thermal-neutralino source (late decays or hidden
    sector dark matter for example)~\cite{leszek}. 
The rest of the analysis proceeds exactly as in section~\ref{sec:Lmaps} (i.e.\
without simultaneously taking theoretical uncertainty into account). 
\begin{table}
\caption{Likelihood of being in a certain region of mSUGRA parameter
  space including possible an additional contribution from
  non thermal-neutralino dark matter. \label{tab:regions3}}  
\begin{tabular}{|c|c|}
\hline Region & likelihood \\ \hline
$h^0$ pole & 0.04$\pm$0.01 \\
$A^0$ pole & 0.52$\pm$0.02 \\
co-annihilation & 0.14$\pm$0.02 \\ \hline
\end{tabular}
\end{table}

The MCMC algorithm turns out to be much more efficient once we drop the
assumption that the cold dark matter consist only of neutralinos:
19.9$\%$ efficiency was achieved compared to 4.1$\%$. One consequence of this
is that statistical fluctuations in the results are smaller. 
The likelihoods in each region are shown in Table~\ref{tab:regions3}. 
A comparison of Tables~\ref{tab:regions},\ref{tab:regions3} shows that
annihilation through $A^0$ pole has acquired a significantly larger likelihood
through allowing for other forms of dark matter. The higgs-pole and
co-annihilation regions are still, within statistics, compatible with their
previous likelihoods. All of the listed uncertainties have decreased, due to
the additional efficiency.

\begin{figure*}
\twographst{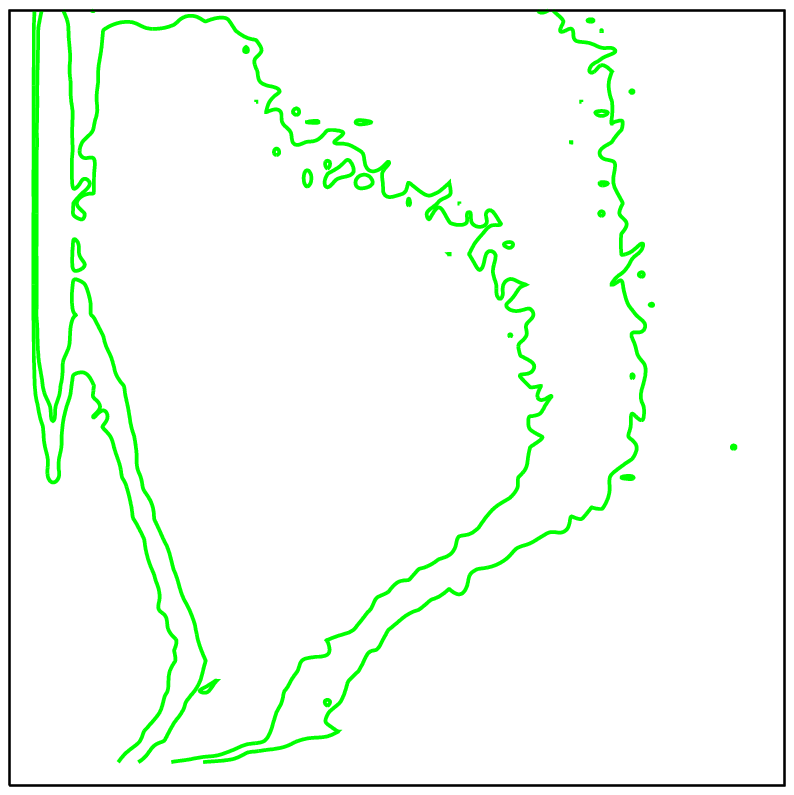}{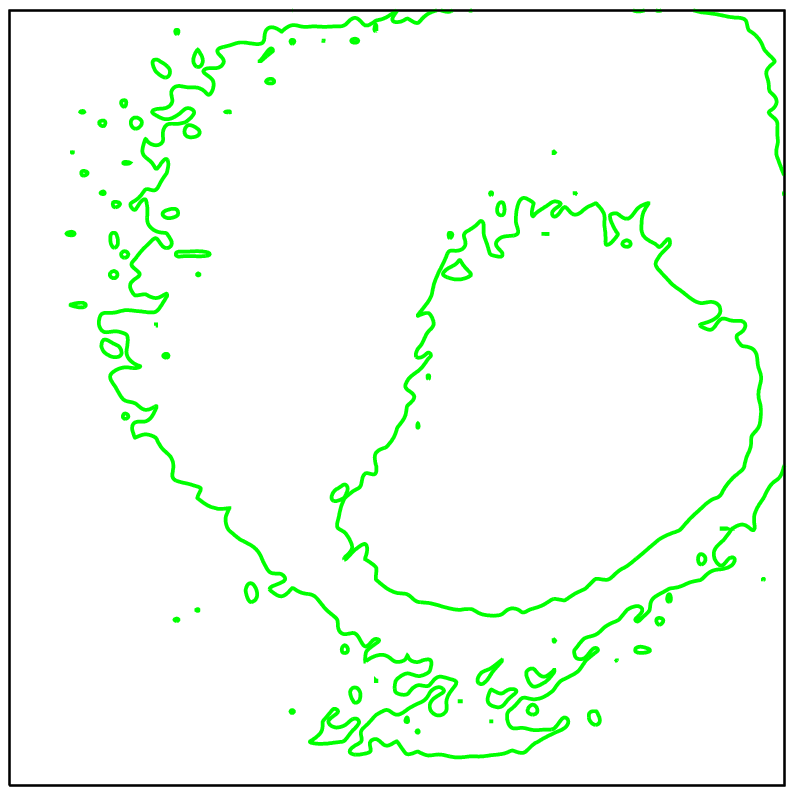}
\caption{Likelihood maps of mSUGRA parameter space allowing for non
  thermal-neutralino contributions to the dark matter relic density. 
The graphs show the
likelihood distributions sampled from
  7d parameter space and marginalised down to two.
The likelihood (relative to the likelihood in the highest bin) 
is displayed by reference to the bar on the right hand side of each plot.
The contours show the 68$\%$ and $95\%$ confidence level limits.
\label{fig:noLow}}
\end{figure*}
Fig.~\ref{fig:noLow}a shows the likelihood map marginalised to the
$M_{1/2}-m_0$ plane. Comparing it to Fig.~\ref{fig:like7}a, we see a very
similar picture except for the fact that the higher volume of likelihood in
the $A^0$-pole region is evident. The same comment can be made of all of the
plots analogous to the ones in to Fig.~\ref{fig:like7}: we show the
likelihood marginalised to the $m_0-A_0$ plane in Fig.~\ref{fig:noLow} as an
example.  There are no other qualitative changes in any of the plots, and
indeed the 
likelihood distributions marginalised to the $A_0-\tan \beta$ and
$M_{1/2}-A_0$ planes (analogous to Figs.~\ref{fig:like7}e,f) 
are identical by eye, except for being smoother due to
the increased efficiency. Likelihoods of sparticle masses also look the same,
except for the spike in the gluino mass, which has twice as much
likelihood. As mentioned before, this spike is due mainly 
to the light $h^0$-pole
region which is subject to relatively large fluctuations, as
Tables~\ref{tab:regions},\ref{tab:regions3} illustrate. We cannot conclude
that the $h^0$-pole region obtains more integrated
likelihood by admitting non thermal-neutralino components to the relic
density because the statistics in the MCMC algorithm are not high enough. 

\begin{figure}
\epsfig{file=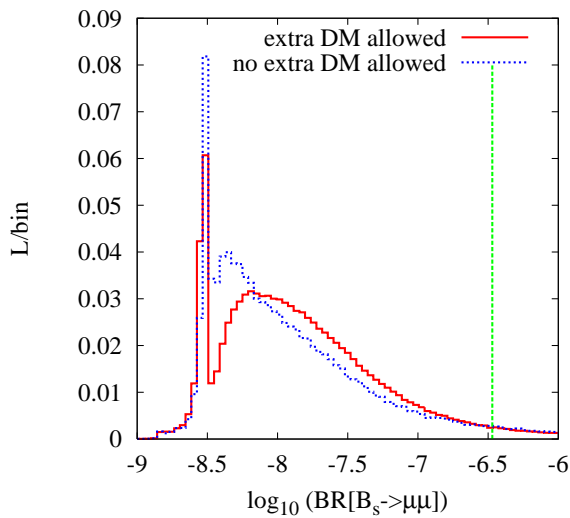, width=\wth}
\caption{Investigation on the effects of allowing for a non
  thermal-neutralino component of dark matter in the 
  branching ratios for the decay $B_s \rightarrow \mu^+
  \mu^-$ in mSUGRA. The current
  Tevatron upper limit is displayed by a vertical line.
\label{fig:BsmumuNo}} 
\end{figure}

One distribution that does significantly change shape is that of $BR(B_s
\rightarrow \mu^+ \mu^-)$. We show its marginalised distribution after
dropping the lower likelihood penalties on $\Omega_{DM} h^2$ from the MCMC
algorithm procedure in Fig.~\ref{fig:BsmumuNo} as the histogram
 marked ``extra DM allowed''. 
For the purpose of comparison, the default calculation where we assume all
dark matter to be thermal neutralinos from 
Fig.~\ref{fig:hists}c is also displayed, being marked ``no extra DM allowed''.
Comparing the two distributions, 
we see a broader distribution due to the enhanced
$A^0$-pole  annihilation region when additional components are allowed in the
fit. The $A^0$-pole region has higher $\tan \beta$, and therefore higher
values for the branching ratio.
The estimated amount of likelihood that could be covered by Tevatron
measurements with 8 fb$^{-1}$ of integrated luminosity increases by 6$\%$ to 
35$\%$ of the currently allowed density due to the presence of non
thermal-neutralino dark matter contributions. 

\section{Conclusions \label{sec:conc}}

Previous studies of mSUGRA in the context of dark matter and particle physics
constraints have tended to not use the full dimensionality of the parameter
space, and to have put hard 95$\%$  (C.L.) limits on
predictions. Here, in the full dimensionality of parameter space, we include
all of the information in a likelihood fit, so that violating one constraint
slightly might be traded against fitting another one better in a consistent
manner. 
Although there is plenty of qualitative information about possible dark matter
annihilation regions in mSUGRA in 
the literature, this paper gives a quantitative calculation of the
likelihood distributions in the full dimensionality of the parameter
space. However, the 
most important contribution of this paper lies in the implications of the
results to particle physics. 

We have successfully employed an MCMC algorithm to provide 
likelihood maps of the full 7d input parameter space of mSUGRA\@. 
By using a statistical test, we have shown that the likelihood distributions
have achieved good convergence before a total of 9$\times 10^6$ samplings of
the likelihood. 
We have presented the likelihood marginalised down to each 2d mSUGRA parameter
pair. Such plots provide the totality of the
current information we have about the model given the experimental
constraints and are quantitative results. Theoretical
uncertainties in the  
sparticle spectrum calculation broaden a couple of the distributions a little
but do not change them radically. 

The main new contribution of this paper is to our knowledge of what current
constraints on mSUGRA mean for particle physics in a quantitative sense. 
Marginalised 1d likelihood distributions of quantities such as sparticle
masses (or mass differences) 
already show some significant structure from the data, providing
interesting information for future collider searches. In particular, the
likelihood of the ``golden cascade'' 
${\tilde q}_L \rightarrow \chi_2^0 \rightarrow {\tilde l}_R
\rightarrow \chi_1^0$ being kinematically allowed is
  24$\pm 4\%$. $m_{{\tilde \tau}_1} - m_{\chi_1^0}$ is peaked below 10 GeV,
implying that stau reconstruction at hadron colliders could be
problematic at the LHC\@. Integrating the likelihood density, we find the
likelihood of $m_{{\tilde \tau}_1} - m_{\chi_1^0}<10$ GeV is 19.9$\%$.
Our likelihood distributions for $BR(B_s \rightarrow \mu^+ \mu^-)$ corroborate
the 
conclusions of ref.~\cite{Ellis:2005sc}: that current bounds upon the
branching ratio do not yet place significant constraints upon mSUGRA once
other constraints have been taken into account, but any improvement on the
upper bounds constrain the currently available parameter space.  
The quantitative results on $BR(B_s \rightarrow \mu^+ \mu^-)$ are particularly
important: if the Tevatron experiments can reach down to 2$\times 10^{-8}$,
they will cover 29$\%$ of the mSUGRA likelihood, or 35$\%$ if we allow the
possibility of additional contributions to the relic density other than 
thermal neutralinos.

We have shown that the correlation between $BR(B_s \rightarrow \mu^+ \mu^-)$
and $\delta 
(g-2)_\mu$ noticed in Ref.~\cite{Dedes:2001fv} is much diluted once
simultaneous variations
of all mSUGRA parameters are taken into account. 
The $A^0$-pole annihilation region
and the stau co-annihilation region each have approximately
an order of magnitude more likelihood than the $h^0$-pole region.
Stop co-annihilation is highly 
disfavoured compared to these other regions due to
more restrictive Higgs mass constraints coming from a lower value of $m_t$, 
as well as the $BR(B_s \rightarrow \mu^+ \mu^-)$ and $\delta 
(g-2)_\mu$ predictions. 
  The light $h^0-$pole region just survives the LEP2 Higgs
  mass constraints despite the new reduced 
  top mass value albeit with a reduced likelihood (in the usual frequentist
  language, it's outside the 95$\%$ confidence level but not the 99$\%$ one).
  The light $h^0$-pole region has more likelihood if one allows additional
  non thermal-neutralino components of dark matter. 

The analysis of
  Ref.~\protect\cite{Ellis:2004tc} 
includes $M_W$ and $\sin^2 \theta_{eff}^l$ in the $\chi^2$ statistic, excluding
$M_{1/2}>1500(600)$ GeV at the 90$\%$ confidence level for $m_t=178$ GeV and
  $\tan \beta=50(10)$ respectively. These numbers are not exactly reproduced
  in the present analysis for several reasons: we use a more up-to-date value
  of $m_t$, we vary $m_t$, $m_b$, $\alpha_s (M_Z)$ and $\tan \beta$
  simultaneously with the other mSUGRA parameters and we do {\em not}\/ include
  $M_W$, $\sin^2 \theta_{eff}^l$ in 
  the fit. Indeed, one may wonder about introducing a posteriori bias as a result
  of only picking these two precision electroweak observables, since they are
  the two that show a preference for a SUSY contribution. Other observables
  would presumably prefer heavier SUSY particles. 
  However, $M_W$ and $\sin^2 \theta_{eff}^l$ do show a preference for lower
  $M_{1/2}$ for $m_t=178$ GeV. Having said that, our results are not wildly
  different, as an examination of Fig.~\ref{fig:like7}a shows. 

We suspect that the MCMC techniques exemplified here could be found 
extremely useful in SUSY fitting programs such as {\tt
  FITTINO}~\cite{Bechtle:2004pc} and {\tt SFITTER}~\cite{Lafaye:2004cn} in
order to provide a likelihood profile of the 
parameter space, including secondary local minima. These programs are designed
to fit more general MSSM models than mSUGRA to data, with an associated
increase in the number of free input parameters, so the linear calculating
time of MCMCs ought to be very useful.

While our results presented for mSUGRA  are in themselves interesting, it is
obvious that the method will be applicable in a much wider range of
circumstances. Once new observables become relevant, such as some LHC
end-points, for example, it would be trivial to include them into the
likelihood and re-perform the MCMC~\cite{whitey}. 
The method should be equally applicable to
other models, and provided enough CPU power is to hand, could provide
likelihood maps for models with even more parameters and/or detailed
electroweak fits.

\acknowledgments
We would like to thank other members of the Cambridge SUSY Working Group,
W de Boer, 
P Gondolo, P H\"afliger, B Heinemann, S Heinemeyer, G Manca, A Peel, M Rauch
and T Plehn 
for helpful conversations. This work has been partially supported by PPARC. 

\appendix
\section{Toy Models and Sampling}
\label{sec:badsampling}

By definition, a sampler able to sample correctly from a pdf
$p({\mathbf x})$ must generate a list of ${\mathbf x}$ values
whose local  
density, at large step-numbers, is proportional to the probability density
$p({\mathbf x})$ at each part of the space (in the preceding parts of the
paper we have set this pdf to be the likelihood).
The value of the constant of proportionality between the probability
density and the local density of ${\mathbf x}$ values is unimportant, but the
key point is that it is {\em constant}\/ across the whole space.

It can sometimes be hard to implement a good sampler for a given
probability distribution.  In fact it is often easier to invent an
algorithm which generates a sequence of ${\mathbf x}$ values, and which may
superficially resemble a sampler, {\em but which lacks constancy of
proportionality over the space}.  We might call such algorithms
pseudo-samplers as their output can sometimes resemble that of a
true sampler, provided that the variation in proportionality is not
too great across the space considered.  Pseudo-samplers are sometimes
useful (e.g.\ as a means of exploring a multidimensional space
in which case sample density may be neither interesting nor the
end product of the analysis).

In the present and in many other papers, however, sample density represents
confidence in some particular part of parameter space, and {\em is}\/
the final product of the analysis.  Extreme care, then, must be taken
to ensure that any creative modifications to established Markov
Chain sampling techniques do not break the principle of
detailed balance -- the test which ensures that the algorithm remains a
true sampler rather than a pseudo-sampler.

It is often desirable for Metropolis-Hastings type samplers to have an
efficiency of about 25\% for the acceptance of newly proposed points.
Efficiencies much smaller than this may suggest that the proposal
distribution is too wide and is too often proposing jumps to
undesirable locations far away from the present point, leading to large
statistical fluctuations in the result.  Efficiencies
much larger than this can be indicative of proposal distributions
which are too narrow and may take too many steps to be practical
to random-walk 
from one side of a region of high probability to the other.
It is tempting, therefore, to adapt the present step size
(i.e.\ proposal distribution width) on the basis of recent efficiency.
With a couple of toy examples, however, we will illustrate that 
this is a dangerous path to follow, and we will demonstrate that it break the
principle of detailed balance, and thus turns the Markov Chain algorithm from a
sampler into a pseudo-sampler.
 
We take as our example the method used by Baltz and
  Gondolo~\cite{Baltz:2004aw} which is 
  designed to keep the target efficiency of the authors' Markov Chain
  close to 25\%:
\begin{enumerate}
\item
Double the current step size if the last
three proposed points were all accepted
\item
Halve the current step size of the last seven proposed points were all rejected.\end{enumerate}
We will refer the the above method as ``the adaptive algorithm'' and show
that it fails to sample correctly from some simple distributions, a signature
of detailed balance being broken.

\begin{figure*}
\twographschris{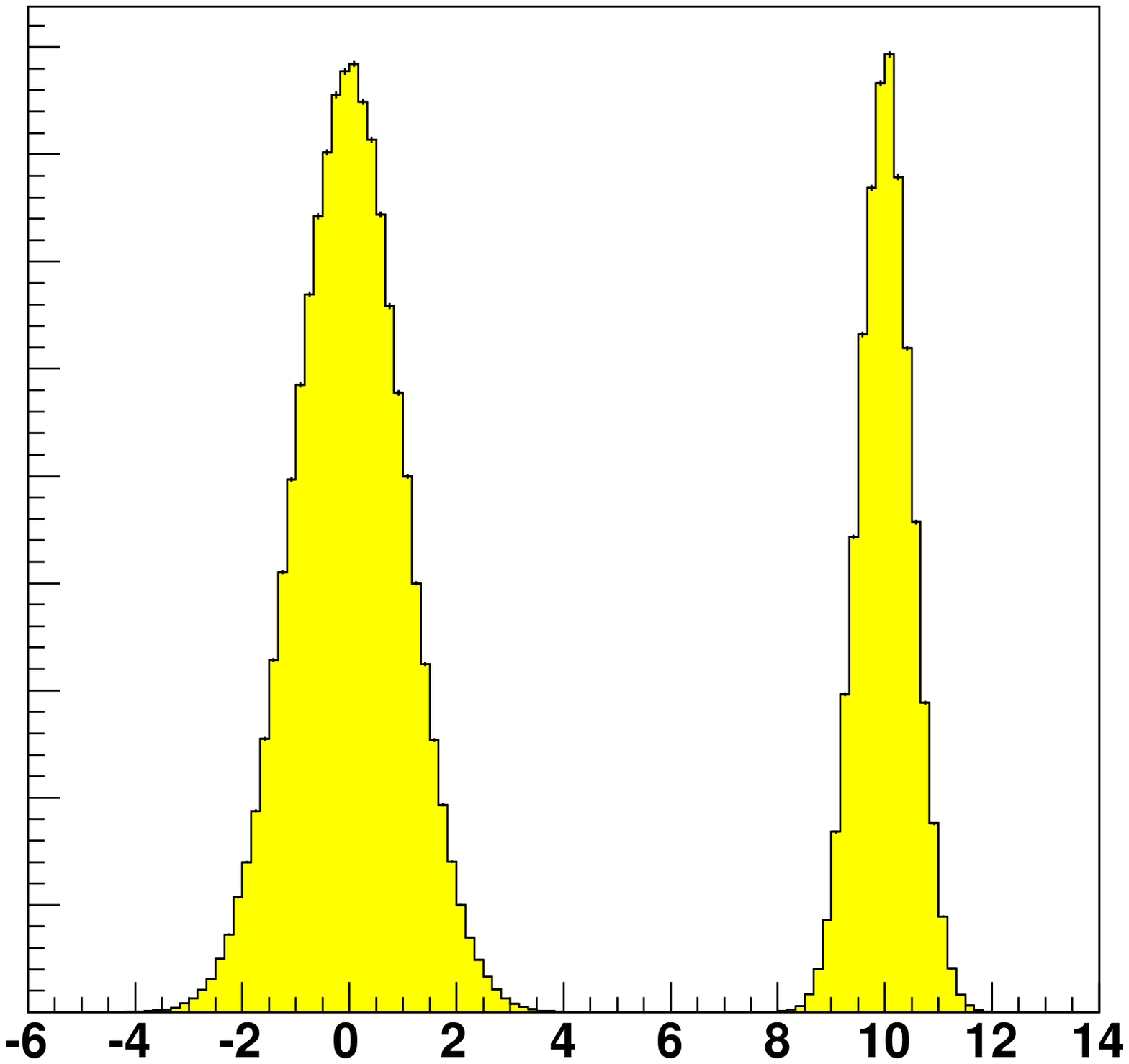}{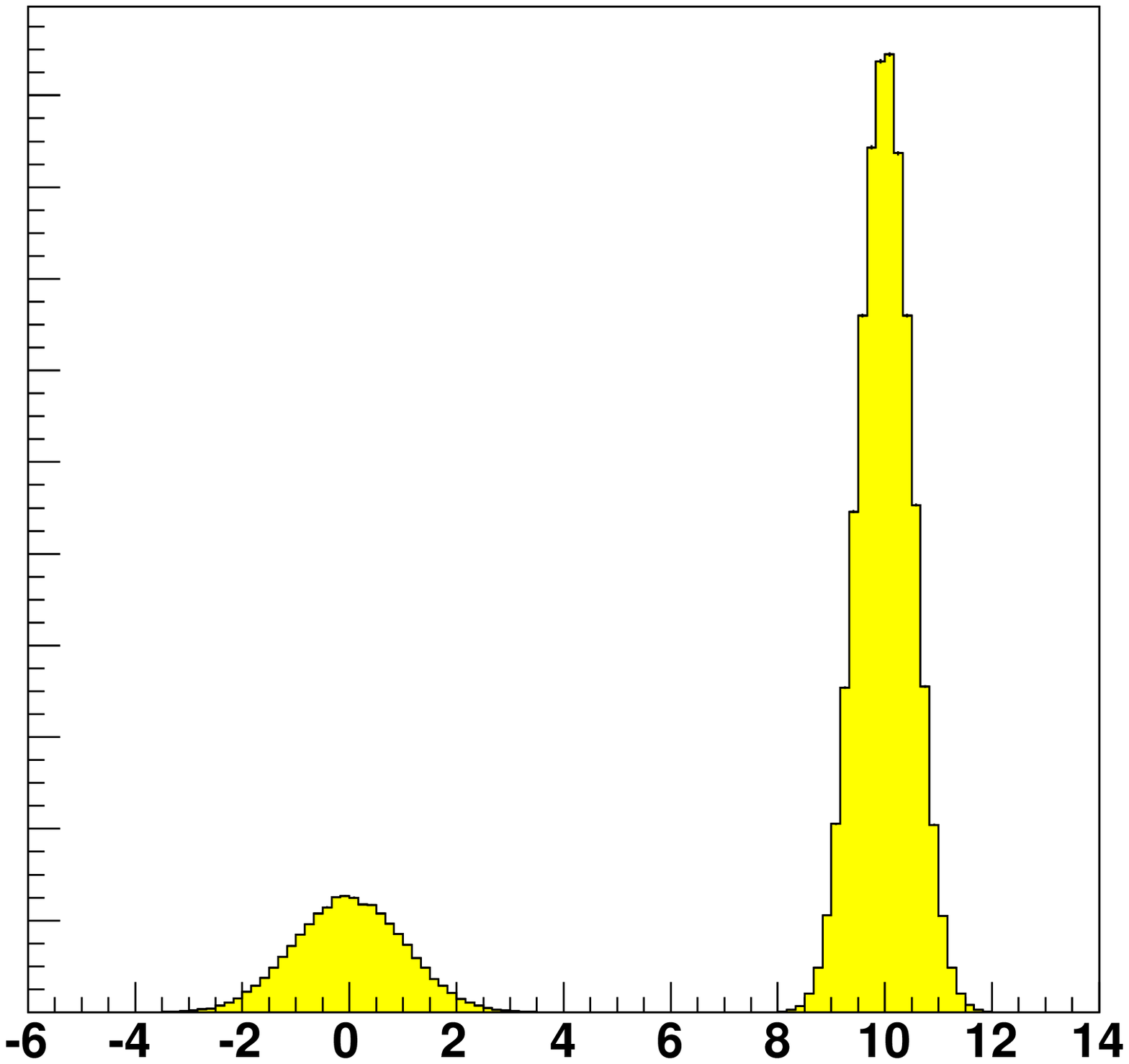}
\caption{Binned samples of the double Gaussian distribution $p_{dGau}(x)$. The
  normalisation is arbitrary and has no relevance   here. 
  (a) uses a Metropolis-Hastings algorithm and yields a good approximation
  whereas (b) uses the adaptive algorithm. 
\label{fig:dGau}}
\end{figure*}
Take, for example, the 1-dimensional double Gaussian probability distribution
$p_{dGau}(x)$ 
defined by
\begin{equation}
p_{dGau}(x) \propto g(x, 0, 1) + g(x, 10, 1/2),\label{eq:dGau}
\end{equation}
where $g(x, m, \sigma)=\exp(-(x-m)^2/(2\sigma^2))$ is a Gaussian distribution
of unit height and width $\sigma$ centred on $x=m$.
Note that the Gaussian near the origin is wider than the
Gaussian near $x=10$.  This pdf mocks up the approximate situation along the
$M_{1/2}$ direction in mSUGRA for large $m_0$, as Fig.~\ref{fig:like7}a
shows. The narrow Gaussian would then correspond to the light $h^0$-pole
region, which is quite disconnected from the wider $A^0$-pole
region. 
A Metropolis-Hastings algorithm (see section~\ref{sec:mcmc})
with a fixed Gaussian proposal pdf $Q(x)$ of width 5 was
run for 2 000 000 steps and the binned result is shown in
Figure~\ref{fig:dGau}a. It reproduces the target distribution very well.
In contrast, Figure~\ref{fig:dGau}b shows what happens when
the adaptive algorithm is run on $p_{dGau}(x)$.  Clearly the result
is very different to that in Figure~\ref{fig:dGau}a and is thus
very wrong.  The narrow Gaussian has been sampled many times more frequently
than it should have been relative to the wider one near the origin.

The adaptive algorithm ensures that whenever the current
point is in one of the two Gaussian regions, the step size is adjusted
to be proportional to the width of that region.  This adaptation is
not immediate (seven successive rejections must occur before the step
size is halved) but suppose for the moment that adaptation were to
take place almost immediately. For the moment, let us also neglect 
random fluctuations of the step size.
In such a limit, when the current
point is in the left-hand Gaussian region, the step size is double what
it is when the current point is in the right-hand region.  Making a
proposal for a jump from the region on the left to the region on the
right, therefore, is something like a $10$-sigma event.  In contrast,
making the reverse proposal (from right to left) is more like a
$20$-sigma event.
In this limit, it is thus $e^{((20)^2 - (10)^2)/2}=e^{150}$
times more likely that 
jumps to the right get proposed than 
jumps to the left.  
This is a vast overestimate of the bias toward
the narrow region for two reasons.  Firstly, step size adaptation is
not immediate (even when the step size is half of what it should be,
quite a few steps occur before
seven successive rejections).  Secondly and more importantly,
even when settled in one of the two regions, the adaptive step size
makes excursions about its mean value.  Excursions to very high step
sizes (double or quadruple the average step size) are infrequent
but still occur.  When they do, they elevate the chance of proposing a
jump from one region to another, and help to equilibrate between the
two regions.  Both of these effects reduce the bias favouring the narrow
region, but still break detailed balance, and overall the right hand
Gaussian is still sampled about 10 times more frequently than it
should be.




\begin{figure*}
\twographschris{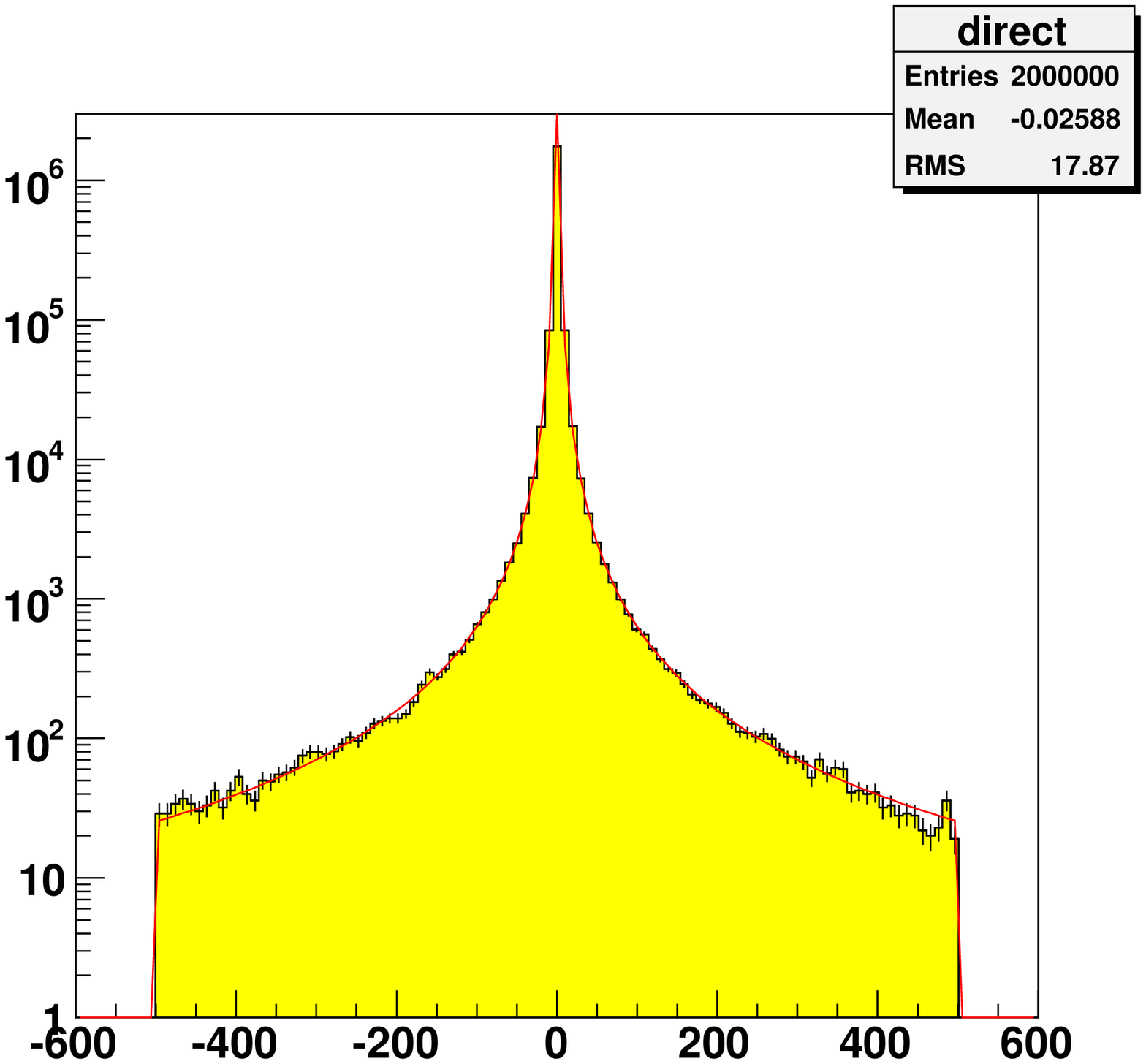}{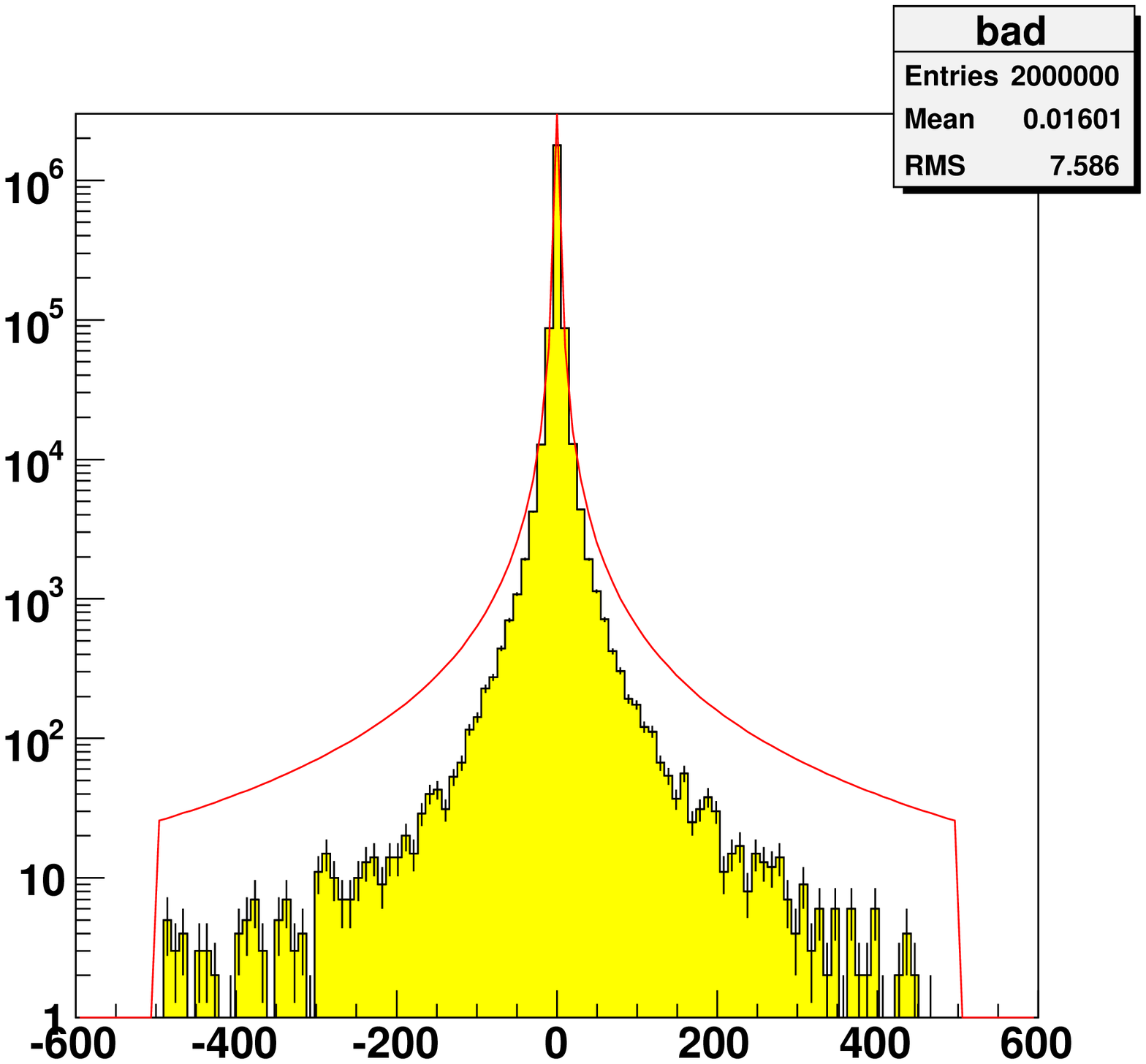}
\caption{Binned sampling of a Cauchy distribution $1/(1+x^2)$ (shown in solid
  red line) truncated at   $x=\pm   500$.
  (a) shows the result of direct sampling and yields a good
  approximation whereas (b) uses the adaptive algorithm. 
\label{fig:cauchy}}
\end{figure*}
For quite a different example, consider the truncated Cauchy
distribution defined by
\begin{eqnarray}
p_{cauchy}(x)\propto
\begin{cases}
1/(1+x^2)& \mbox {if~}x>-500 \mbox{~and~} x<500, \\
0 & \mbox{otherwise}.
\end{cases}
\label{eq:cauchydist}
\end{eqnarray}
A faithful sampling from this distribution is shown in
Figure~\ref{fig:cauchy}(a), and a mis-sampling using the adaptive
  algorithm is shown for comparison in Figure~\ref{fig:cauchy}(b).
Although there is better correspondence between the samples
generated by the two methods than was the case earlier, it is
nevertheless evident that there are large differences between the
degrees to which the two methods have sampled the tails of the
distribution.  As a consequence, samples from the adaptive method
have a root-mean-squared value of 7.6 compared to the value of 17.9 obtained
by the true 
sampling.  The cause of the discrepancy is again due to the very
different scales in the cauchy distribution.  Its core is very narrow
with a width of order 1, but there are significant parts of probability also
lodged in the tails many orders of magnitude away.  The adaptive
method has to raise and lower the step size frequently to sample from
the whole dynamic range, and in doing so it has to break detailed
balance many times, resulting in the cumulative effect of an overall
order of magnitude bias in the tails.

\providecommand{\href}[2]{#2}\begingroup\raggedright\endgroup


\end{document}